\documentclass[a4paper,twoside,11pt]{article}

\usepackage[USenglish]{babel} 
\usepackage[T1]{fontenc}

\usepackage{graphicx}
\usepackage[labelsep=endash, figurename=Figure, tablename=Table, textfont=it]{caption}
\usepackage{subcaption}
\usepackage{float}  
\usepackage{amsmath}
\usepackage{hyperref, wasysym}
\usepackage{multicol}
\usepackage{amsmath}
\usepackage{enumitem} 
\usepackage{booktabs}
\usepackage{amssymb}
\usepackage{array}
\usepackage{pifont} 

\usepackage{fancyhdr}
\usepackage[top=2.5cm, bottom=2.5cm, left=2.5cm, right=2.5cm]{geometry} 
\usepackage[usenames,dvipsnames]{pstricks}
\usepackage[round,comma,authoryear]{natbib}
\usepackage{titling}
\usepackage{longtable}
\usepackage{pdflscape} 
\usepackage{enumitem,amssymb} 
\usepackage{multirow} 
\newlist{todolist}{itemize}{2} 
\setlist[todolist]{label=$\square$} 
\newcommand{\checkeditem}{\item[\checkmark]} 
\newcommand{\naitem}{\item[---]} 
\newcommand{\xmark}{\item[\ding{53}]} 

\definecolor{darkblue}{rgb}{0,0,0.5}
\hypersetup{colorlinks=true,linkcolor=black,citecolor=darkblue,urlcolor=darkblue}

\usepackage{xcolor}

\newcommand{\yes}{\textcolor{Green}{\ding{52}}}
\newcommand{\soso}{\textcolor{Dandelion}{\ding{108}}}
\newcommand{\no}{\textcolor{BrickRed}{\ding{56}}}

\makeatletter
\let\ps@plain\ps@fancy
\makeatother

\begin{document}


\title{\bf The Calibration Illusion in Traffic Microsimulation}

\author{Cameron Hickert$^{1, 2, *}$, Maryam Samaei$^{3}$, Athena Wang$^{4}$, Chengyuan Zhang$^{5}$, \\ Lijun Sun$^{5}$, Yanbing Wang$^{6}$, Mostafa Ameli$^{3, 7}$, and Cathy Wu$^{1, 2, 8}$}
\date{August 18, 2026}

\pretitle{\centering\Large}
\posttitle{\par\vspace{5ex}}

\preauthor{\centering\large}
\postauthor{\par\vspace{2ex}
\small
$^{1}$ MIT Laboratory for Information \& Decision Systems, Cambridge, USA\\
$^{2}$ MIT Institute for Data, Systems, and Society, Cambridge, USA\\
$^{3}$ GRETTIA, COSYS, Univ. Gustave Eiffel, Paris, France\\
$^{4}$ MIT Dept. of Electrical Engineering \& Computer Science, Cambridge, USA\\
$^{5}$ Dept. of Civil Engineering, McGill University, Montreal, Canada\\
$^{6}$ School of Sustainable Engineering and the Built Environment, Arizona State University, Tempe, USA\\
$^{7}$ Department of Electrical Engineering and Computer Sciences, University of California, Berkeley, Berkeley, USA\\
$^{8}$ MIT Dept. of Civil \& Env. Engineering, Cambridge, USA\\
\vspace{1ex}
$^{*}$ Corresponding author: chickert@mit.edu

\vspace{2ex}\it
Working Paper
\vspace{1ex}

}

\maketitle

\begin{abstract}
The transportation community seeks to use calibration methods for highway traffic microsimulation. This is a response to the time-consuming and subjective nature of traditional manual calibration, as well as the growing prevalence of data for calibration. This work argues that this ``automatic'' calibration is largely an illusion. A significant – and unquantified – amount of bespoke manual work is hidden behind these methods. This illusion inhibits a core component of scientific advancement: objective comparison against a shared standard. This impedes evaluation, undermines reproducibility, and fragments research. To address this gap, this paper introduces a comprehensive benchmark designed to simultaneously expose the calibration illusion for highway microsimulation and provide a common ruler. The results across a range of scenarios present a new baseline for what the algorithms can achieve without bespoke tuning, revealing the research gap that remains and providing a tool to advance a cumulative science of calibration. Additional experiments provide insights into the source of calibration errors that arise in large-scale highway calibration relative to the simplified settings under which methods are commonly developed.

\end{abstract}

\noindent\rule{\textwidth}{0.5pt}\vspace{0cm}

\fancypagestyle{firststyle}{
\lhead[]{}
\rhead[]{}
\rfoot[]{}
\cfoot[]{}
}
\thispagestyle{firststyle}

\pagestyle{fancy}
\fancyhead{}
\fancyfoot{}
\renewcommand{\headrulewidth}{0pt}
\renewcommand{\footrulewidth}{0pt}
\setlength{\headheight}{15pt}
\rhead[\thepage]{\thepage}
\rfoot[Working Paper]{Working Paper}
\cfoot[]{}


\newpage

\tableofcontents

\newpage

\section{ INTRODUCTION}
\label{sec:introduction}

The goal of traffic microsimulation calibration is to systematically adjust the parameters of a microscopic traffic simulation model so that the virtual environment accurately reproduces empirical traffic dynamics. By systematically adjusting parameters -- such as car-following model elements or lane-changing behaviors -- researchers seek to minimize discrepancies between simulated outcomes (e.g., flow rates, vehicle speeds, travel times) and empirical measurements collected from actual roads.

A tuned model is a necessary foundation for downstream evaluation, such as assessing congestion mitigation strategies or testing emerging traffic control methods [\cite{dowling2004guidelines}]. In essence, calibration provides a strong basis for both academic research and practical traffic management decisions by matching the simulation to real-world conditions.

Unfortunately, manual calibration is time-consuming and labor-intensive. One study assessing the labor demands of freeway microsimulations found data collection, model development, and calibration for a single environment typically consume between 180 and 4,500 working hours [\cite{alexiadis2014guidance}]. In response to this -- as well as the subjective nature of traditional manual calibration and the growing prevalence of data for calibration -- the transportation community seeks to replace manual tuning with automated calibration methods for highway traffic microsimulation, such as genetic algorithms [\cite{ciuffo2013no}]. These present an opportunity to bypass or accelerate the time-consuming manual process and allow application beyond a specific road scenario in which a researcher or transportation engineer has local expertise. 

This work argues that such ``automatic'' calibration is largely an illusion. We find that a significant -- and unquantified -- amount of bespoke manual work is hidden behind these methods. This includes the selection of parameter ranges, manual definition or exclusion of parameters deemed unimportant, and the use of local expertise to tune initial conditions. These manual interventions affect performance and thus are components of the true method, but are rarely documented.

This illusion presents three challenges to a cornerstone of scientific advancement: objective comparison against a shared standard. First, it impedes evaluation, as it is unknown to which extent a result is due to an algorithm's independent merit or a researcher's unreported manual tuning. Where researchers may believe they are comparing automated methods, in effect they may be comparing ambiguous combinations of automated methods and human expertise. Second, because the methods as-published are incomplete, it compromises reproducibility and thus inhibits advancement in the field. In cases where multiple solutions are permissible, reproducing prior work can require more effort than tuning required in the original work, as researchers search among various feasible solutions for the specific one previously described. Finally, the barriers created discourage researchers and practitioners from adopting new methods and settings, siloing research efforts. It is often more practical to repurpose a method or setting from one's own previous work or that of a direct collaborator, rather than to adopt the state-of-the-art. Even if methods were fully specified, the absence of a shared benchmark inhibits a cumulative science in this domain.

Traffic microsimulation is a critical stage in the design and analysis of traffic systems, yet the credibility of these studies hinges on accurate calibration. 
While the community has produced various calibration algorithms -- and even benchmarked \textit{macro}scopic traffic models [\cite{mohammadian2021performance}] -- a common benchmark for \textit{micro}scopic simulations remains absent.
Thus the need remains for a mechanism to systematically evaluate and compare highway traffic calibration performance across scenarios and parameter types (e.g., car-following, lane-changing, and origin-destination).

To resolve these issues, this paper offers \textbf{three primary contributions}. 
\textbf{First}, it introduces AutoTune, a comprehensive benchmark designed for automatic calibration of highway traffic microsimulation across networks of varying spatial and temporal scales (small, medium, and large). Expanding beyond preliminary frameworks, this environment includes both synthetic and real-world aggregate and trajectory data. This includes data, open-source simulation and calibration code, and a framework for evaluation, each of which the community can continue to utilize as a living benchmark that enables objective measurement and comparison of fully-specified, truly automatic methods against a common standard. As the transportation community continues to invest in data collection, this benchmark can provide a missing link to develop the rigorous calibration methods needed to translate that data into reliable models. \textbf{Second}, we characterize the calibration illusion, leveraging the benchmark's ``glass box'' design that aids transparency by including both parameters traditionally considered in automatic calibration, as well as those normally excluded or tuned manually. For each approach evaluated, the open-source code is the method in its entirety. \textbf{Third}, we present results across small, medium, and large-scale traffic settings that present a new baseline for a cumulative science of calibration, detailing what algorithmic calibration can (and currently cannot) achieve without bespoke tuning, thereby revealing the gap that remains. Additional experiments provide insights into the source of calibration errors that arise in large-scale highway calibration relative to the simplified settings under which methods are commonly developed. 

The \textit{AutoTune} name combines \textit{auto}mobiles with the concept of parameter \textit{tuning}, while also emphasizing the framework's mission: advancing fully \textit{auto}mated highway simulation calibration.
For many users, the objective will not be a method's performance on the specific networks used in the simulation, but rather the fair and open development and evaluation of methods that can then be applied to a scenario of specific interest to them. AutoTune thus offers a platform for more transparent, reproducible, and comparable research outcomes in this pivotal area of transportation science.

\subsection{Paper Structure}
The structure of the paper is as follows. Section~\ref{sec:related_work} reviews the relevant literature. Necessary background and notation is introduced in Section~\ref{sec:background_notation_preliminaries}, ahead of Section~\ref{sec:problem}, which formulates the problem. Section~\ref{sec:methodology} describes our methodology in constructing the benchmark and details the methods assessed. Experiments leveraging the benchmark to assess the performance of automatic methods and characterize the calibration illusion are detailed in Section~\ref{sec:experiments}. Section~\ref{sec:results} presents the experimental results and associated discussions. Finally, Section~\ref{sec:conclusion} concludes.

\subsection{Prior Work and Disclosures}
This manuscript is an extended and substantially revised version of preliminary work presented at the IEEE Intelligent Vehicles (IV) Symposium [\cite{hickert2026autotune}]. While the initial conference proceeding introduced the foundational AutoTune framework and a single proof-of-concept case study on a single network, this article significantly expands the scope and depth of the research. Specifically, this work formally introduces and characterizes the ``calibration illusion,'' broadens the experimental evaluation to encompass two additional networks and all-new scenarios on the large network with explicit noise evaluations, and provides extensive ablation studies quantifying the impact of traditionally excluded parameters. Additionally, in the interest of transparency, the authors acknowledge the use of large language models (LLMs) for drafting assistance and copyediting, as well as for coding assistance during the development of the benchmark's software infrastructure. The authors have reviewed and edited all AI-generated output and take full ownership and responsibility for the final content of this publication.

\section{ RELATED WORK}
\label{sec:related_work}

\subsection{Microsimulation Calibration}

\subsubsection{Related benchmarks, reviews, and studies}
Over the last 20 years, a growing body of research has focused on calibration methods. A limited, parallel body of comparison and benchmarking efforts -- including work more rigorously evaluating appropriate calibration methodologies -- has more recently taken shape. Our work continues this line of research.

While benchmarking work has been established for origin-destination flow calibration [\cite{antoniou2016towards, rong2024large}] and \textit{macro}simulation continuum models [\cite{mohammadian2021performance}], these frameworks do not translate to microsimulation. By their nature, macrosimulators fail to capture vehicle-level dynamics [\cite{sekar2023micro}], and the lack of open-source availability in existing macroscopic benchmarks further limits community-driven advancements [\cite{mohammadian2021performance}].

We are only aware of two related open-source microsimulation benchmarking efforts: \cite{punzo2021calibration} and \cite{montali2024waymo}. However, the former considers only CF parameters and relies on data from relatively few vehicle trajectories (39 total). The latter -- although large-scale -- is geared toward autonomous driving and thus comprises very short scenarios (9-20 seconds) that preempt OD specification and prevent the analysis of aggregate traffic phenomena of interest. 

Only two open-source benchmarking projects currently exist in the microsimulation space to our knowledge [\cite{punzo2021calibration, montali2024waymo}]. However, both possess structural limitations that prevent holistic calibration. The first focuses exclusively on car-following parameters using a small sample of 39 trajectories. The second targets autonomous vehicle agents using short 9- to 20-second clips. These time horizons prevent the specification of origin-destination flows or the observation of macroscopic traffic patterns over time. Consequently, the transportation community still lacks an open, large-scale framework capable of assessing the accuracy of long-term simulation across a spectrum of behavioral and network parameters.

See Table~\ref{tab:related_benchmarks} in Appendix~\ref{appendix:related_work} for an overview and comparison of related benchmarks, reviews, and studies. It assesses and describes 16 separate attributes for each work. We observe that car-following calibration is the most commonly benchmarked parameter category. However, even among these works, only one [\cite{punzo2021calibration}] open-sources both its data and calibration code. Even \cite{chen2023follownet}, which aims to explicitly benchmark car-following behavior modeling, only releases data and parameters output from its genetic algorithm calibration method, but does not release the calibration methods nor provides sufficient algorithm details to enable full reproducibility. Furthermore, we identified that simulation parameters are rarely considered. 

\subsubsection{Calibration Methodologies and Design Practices}

\begin{figure}[tbp]
    \centering \includegraphics[width=0.95\textwidth]{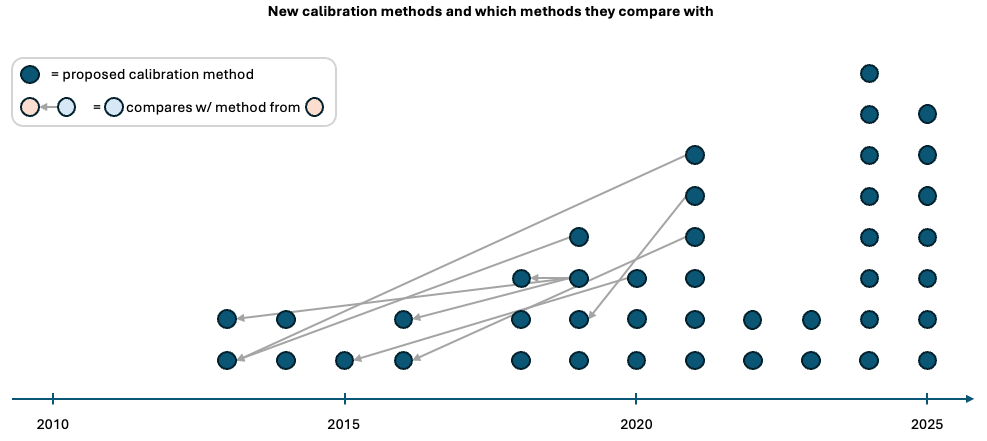} 
    \caption{A timeline of various recently proposed novel microsimulation calibration methods and those with which they compare (if any).}
    \label{fig:new_methods_timeline-blue} 
\end{figure}

Transportation researchers introduce novel calibration techniques, operating on the premise that these algorithms advance the state-of-the-art. As shown in Figure~\ref{fig:new_methods_timeline-blue}, for microsimulation calibration this foundational assumption is on shaky ground. The figure demonstrates how few new methods rarely compare with previous ones. Even those that compare tend to do so against methods which themselves are not compared to others, creating isolated pockets of comparisons. Of the analyzed works, only one paper compares with more than one prior method.

\begin{figure}[tbp]
    \centering \includegraphics[width=0.95\textwidth]{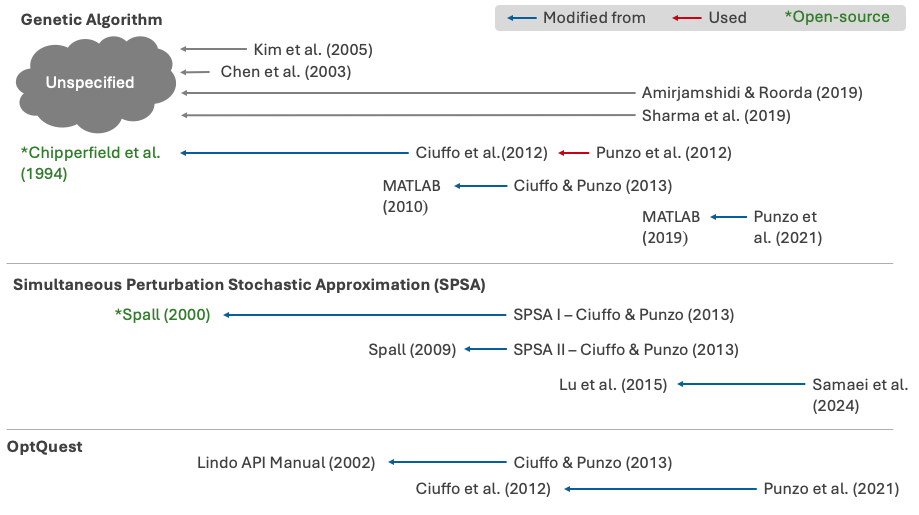} 
    \caption{A history of implementations of popular methods. Except for the work indicated with the single red arrow, each work uses a separate version of each optimization method. Grey arrows also represent potential wasted time from re-implementing methods from scratch.}
    \label{fig:pop_methods_timeline} 
\end{figure}

Even among works that use the `same' method, performance is sensitive to differences in implementations (incl. selection of algorithm hyperparameters, simulation parameters being calibrated, etc.). Figure~\ref{fig:pop_methods_timeline} details the lineages for implementations of popular calibration method in relevant microsimulation calibration benchmarks and comparative studies. Notably, even these well-known methods lack canonical implementations, inhibiting the development of cumulative knowledge as separate versions cannot be assumed to have equal performance. Except for one work, each paper uses a separate version of each optimization method. This limits the community's ability for rigorous comparison and the extraction of meaningful insights. The figure not only highlights the siloing of research work but also indicates wasted effort for researchers as implementation work is rarely reused. 

While calibration methods vary, our review found versions of the genetic algorithm (GA) to be particularly popular, included in six works [\cite{kim2005calibration}, \cite{punzo2021calibration}, \cite{amirjamshidi2019multi}, \cite{chen2023follownet}, \cite{punzo2012can}, \cite{ciuffo2013no}]. We thus include it in our benchmark, along with simultaneous perturbation stochastic approximation (SPSA), another popular method [\cite{ciuffo2013no}, \cite{samaei2024integrating}]. While \cite{ciuffo2013no} conclude that for their problem setting the OptQuest Multistart algorithm performed the best, it is proprietary software, and thus in the interest of transparency and open-source reproducibility we have not included it in our analysis.

A range of data sources, measures of performance, and goodness-of-fit functions are used, which we believe supports the need for a standard benchmark to assess the relative value of various configurations of these. See Table~\ref{tab:related_benchmarks} for further details. The data leveraged in this benchmark differs from that used in previous work in temporal, spatial, and numerical quantity, as well as two qualitative dimensions. More specifically, AutoTune incorporates 10 days each with 4 or more hours of time-continuous data (temporal) across 6.75km of two-way road [spatial] including $\sim$600,000 vehicles [numerical]. Qualitatively, our benchmark includes data at both microscopic \textit{and} macroscopic levels for (nearly) all vehicles in the setting (as opposed to the use of probe vehicles, which generally represent a small fraction of roadway participants [\cite{amirjamshidi2019multi}]).

Conceptually, this study builds upon the findings of \cite{ciuffo2013no}, who demonstrated that the `No Free Lunch' theorem applies to traffic model calibration. This indicates that no single optimization algorithm will prevail in every context. Their research highlighted critical methodological practices that have informed various architectural choices of our benchmark, detailed in the following sections. These authors use synthetic data to investigate how CFM calibration methods recover parameters. However, as (1) it may be the case that no model can capture real-world behavior, and (2) researchers and practitioners often apply pre-designed calibration models to real-world data for which no `ground truth' is available, our benchmark relies on non-synthetic data. Other notable differences are outlined in Table~\ref{tab:related_benchmarks}.

\cite{ciuffo2013no} also highlight the need for further work to help identify classes of problems that may be addressed similarly, as well as those that may not. This suggests a need for benchmarking initiatives and open calibration challenges in which researchers can apply different methods to common case studies. A related gap lies in data availability and quality. It is important to understand the value of having (or not having) various data types. While trajectory data is powerful, not all regions or projects have access to it, and using only aggregate count data can limit calibration fidelity. Bridging this gap might involve developing methods that can calibrate with sparse or partial data, or transfer learning approaches that leverage calibrated models from data-rich areas to data-poor ones.

As the transportation field moves towards calibrations with more realism, expanded scope, and enhanced capabilities, it is important to have a benchmark which incorporates both microscopic and macroscopic data and a fuller range of parameters that affect them. For example, one may want to calibrate not just the average flow, but also the distribution of vehicle headways and lane-changing aggressiveness. This leads to simulations that not only match macroscopic measures but are more realistic on a microscopic level (important for applications like safety analysis or connected autonomous vehicle testing). And because these inter-vehicle and traffic-level behaviors are simultaneously affected by multiple parameter categories -- since, for example, desired headways may be both a function of driver preference and the overall rate of traffic inflow and outflow -- it is important to have a unified benchmark for creating, evaluating, and sharing calibration approaches.

\subsection{Highway calibration}

Two particularly relevant highway traffic calibration works are those by \cite{samaei2024integrating} and \cite{wang2024calibrate}, which also focus on the I-24 MOTION segment of interstate. The former proposes a bi-level calibration framework using probe vehicle data and Tennessee Dept. of Transportation (TDoT) detector measurements. The latter uses a SUMO simulation-in-the-loop approach on a westbound segment of the I-24 corridor using TDoT detector data. We include these methods in our benchmark as examples of the state-of-the-art, given their recency and relevance. Further details are described in \ref{subsec:benchmarked_methods}.

These works highlight the community's interest in I-24 MOTION as a testbed for calibration methods, modeling, and simulation. The results in \cite{samaei2024integrating} were used in preparation for the MegaVanderTest, an experiment on the MOTION stretch of roadway which deployed 100 controlled vehicles to investigate their impact on traffic [\cite{ameli2024designing}]. Given the quality of MOTION's sensing infrastructure, we expect researchers will continue to target it and thus may benefit from the present work.

\section{ BACKGROUND, NOTATION, AND PRELIMINARIES}
\label{sec:background_notation_preliminaries}

\subsection{Concepts}

\subsubsection{Traffic observations}
\label{subsec:traffic_obs}

Define $X = \{x_0, x_1, \dots, x_T\}$ as the sequence of true vehicular trajectory states recorded across a specific time horizon $T$. Because high-fidelity trajectory data is often difficult or expensive to obtain at scale, researchers seldom have direct access to the complete set $X$. Consequently, calibration often relies on a macroscopic set of observed data, defined as $\mathcal{O} = h(X)$, with $h$ acting as the aggregation function. This function $h$ represents the temporal or spatial grouping regularly performed by standard roadway sensors, such as radar systems, induction loops, or connected-vehicle probe streams. Typical macroscopic metrics within $\mathcal{O}$ feature lane-specific flow rates or mean velocities captured at fixed intervals. Such datasets are available for highways in Africa, Asia, Europe, North America, and South America [\cite{simbeye2022deployment, gonzales2009multimodal, fu2020empirical, geroliminis2008existence, cherrett2000traffic, soriguera2011estimation, grote2018practical, middletonvehicle, balid2018real, zechin2020influence}]. The PeMS data mentioned above is one such example. Today this data is cheaper and easier to collect than comprehensive trajectory data, making methods that only require aggregate data more scalable.

\subsubsection{Traffic microsimulation}
\label{subsec:traffic_microsim}

Let $\Theta$ represent the entire parameter space of a given microsimulator, comprising both continuous and discrete variables. We denote the traffic dynamics model as $F_\theta$, which generates the subsequent system state via $x_{t+1} = F_\theta(x_t, u_t)$. In this mapping, $\theta \in \Theta$ defines the set of selected simulation parameters, $x_t$ represents the current traffic state, and $u_t$ denotes control inputs at step $t$. The resulting chronological trajectory of states generated by the model is thus defined as $\tilde{X}_\theta = \{\tilde{x}_0, \tilde{x}_1, \dots, \tilde{x}_T\}$. Applying the aggregation operator to this simulated output yields the corresponding simulation-generated macroscopic dataset, $\tilde{\mathcal{O}}_\theta = h(\tilde{X}_\theta)$.

\subsubsection{Parameter categories}
\label{subsec:param_categories}

Traditionally, elements in $\theta$ are subdivided into three categories: origin-destination parameters, car-following model parameters, and lane-change model parameters. We consider each of these types and add two additional subcategories which are less commonly made explicit: simulation parameters and heterogeneity parameters. Descriptions of each parameter set are below. 

\paragraph{Origin-destination (OD) parameters} As indicated by their name, OD matrices specify origins and destinations for vehicles in a microsimulation over a fixed period. OD parameter sets can thus include either the individuals entries in OD matrices themselves or the parameters used for algorithms to produce those matrices. By allowing various methods to both select and set these parameters, our benchmark can accommodate a wide range of OD calibration methods.

A related but separate concept is that of dynamic traffic assignment (DTA), which is often defined to include vehicle routing and the timing by which vehicles enter into the road network. Our benchmark does not address routing separately due to the network's simplicity. Edge-level routes are fully specified by OD pairs. Lane-level changes are handled by a lane-change model. Furthermore, timing is addressed implicitly by the temporal segmentation of OD flows (for example, by using multiple OD matrices to describe network demand over a single time period) and by the selection of simulation length. 

OD matrix calibration is well-studied and particularly challenging on complex networks with many possible origins and destinations. Given the simplicity of our network, we do not intend our benchmark to quantify OD performance on large-scale networks. Rather, we believe the benchmark's value in this domain is to allow for the co-optimization of OD parameters in parallel to other parameters which jointly define traffic phenomena of interest. This is well-suited to the highway setting, since at moderate timescales and distances, many highways can be characterized with few origins and destinations relative to urban settings. 

\paragraph{Car-following model (CFM) parameters} CFMs are ordinary differential equations that represent how one vehicle follows another on a roadway and are well-studied, including through the use of benchmarks. Popular CFM examples include the Krauss Model and the Intelligent Driver Model (IDM). Parameters for these models may include both those with physical interpretations (such as desired freeflow velocity, desired time headway) and those without. CFM calibrations often rely on trajectory-level data [\cite{chen2023follownet}].

\paragraph{Lane-change (LC) model parameters} LC models are similar to CFMs, but instead determine lane-change behavior. LC parameterization has been studied both separately from CFM parameterization as well as in combination with it, although it is the focus of fewer works in comparison to CFMs.

\paragraph{Simulation parameters} All microsimulations include additional parameters that are artifacts of the simulator itself. A simple example is the simulation's time discretization (timestep length), which interacts with the CFM and LC by setting the frequency at which vehicles may take actions. A more complex example is the warm-up time -- the time allowed for vehicles to flow into the empty traffic network to achieve realistic steady-state traffic density. On the one hand, allowing longer warm-up times may better achieve a desired initial state for the simulation; on the other hand, longer warm-ups may also allow errors to compound. 

These parameters are crucial for reproducibility. Unfortunately, they are rarely mentioned in traffic microsimulation works, and it is even more uncommon that these parameters are made explicit and their calibration methods described. We hypothesize that these parameters are often forgotten or manually calibrated through trial and error. Not only can these parameters affect simulation performance, but they can also impact the computation time of the simulation. This is of particular importance both as the research community seeks (i) to grow the scale of traffic microsimulation and (ii) to enable the use of more data-intensive machine learning methods. 

\paragraph{Heterogeneity parameters} A growing number of works in the field are accounting for traffic heterogeneity, often as part of car-following or lane-change models. Additionally, heterogeneity in vehicle attributes like type or emissions profile may be important for reducing simulator error on metrics of interest. As such, we highlight this parameter set here due to it's cross-cutting nature.

\subsubsection{Joint consideration of parameters}

CFM and LC model calibrations are often handled separately from each other, except in the case of a few notable exceptions. To the best of the authors' knowledge, no works jointly address OD calibration along with CFM or LC calibration. In part this is because OD calibration is often done with aggregate (macroscopic) data, while parameters for CFMs and LC models are often calibrated with trajectory-level (microscopic) datasets. Note, however, that microsimulations common in the transportation community \textbf{require} specification of parameters from \textbf{all} of the above categories, whether or not they are made explicit. For example, a microsimulation work focusing on OD calibration must still utilize a CFM, and conversely, the vehicles in a CFM-focused simulation must have some origin and destination. This underscores the importance of joint consideration of parameter categories. 

This work treats parameters jointly for three reasons. First, if we want to calibrate traffic microsimulations for real-world highways, we often don't have access to trajectory-level data. Thus, one line of effort on this benchmark is to better understand the performance gap between parameters calibrated purely on aggregated observation data and those calibrated on both aggregate and trajectory-level data. For researchers and practitioners alike, this can help answer the question: Is it cost-effective to gather trajectory data? 

Second, each parameter set affects the other. For example, it is known that both traffic volume and aggressive car-following driving styles can contribute to traffic congestion. In a microsimulation with insufficient congestion along a road, it may not be clear whether the CFM parameters or the OD parameters need adjustment. Conversely, error in one may produce error in the other, for example if an artificially aggressive CFM parameterization partially compensates for an erroneous OD matrix or if an OD matrix parameterizes artificially low vehicle demand to partially compensate for an erroneous CFM model. Indeed, especially for larger and longer simulations, we expect that such interaction effects will grow in both frequency and importance due to the greater error compounding that may occur. These interactions mean that we expect separate, sequential tuning of parameters can (and often will) result in different parameterizations and resulting simulation qualities. 

Third, even if a true parameter hierarchy exists that permits sequential calibration, it is currently unknown. By imposing no assumptions about parameter groupings or structure, the benchmark allows researchers to experiment with their own formulations and may accommadate new developments in the field as they occur. This extends to the parameters themselves. For example, instead of using time-continuous CFM models to guide one could choose to benchmark cellular automaton models or artificial neural network models, each of which has their own parameters. This is why we consider traffic microsimulation calibration to consist not only of adjusting chosen parameters, but also of choosing which parameters to adjust. 

\subsection{Traffic microsimulation calibration}
\label{subsec:traffic_microsim_cal}

Given a parameter space $\Theta$, the goal of calibration is to find the parameters $\theta \in \Theta$ that minimize the discrepancy between the traffic simulation model and observations from actual roads. Thus, we can represent a calibration method as  $c: X \times \mathcal{O} \to \Theta$. Note that the calibration methods $c$ of interest in this paper may take as an argument $X$, $\mathcal{O}$, or both $X$ and $\mathcal{O}$. This will aid in understanding the trade-offs of both the richer trajectory-level (microscopic) data in comparison to the aggregate (macroscopic) data, particularly given the costs of collecting the former.

Traditionally, microsimulation calibration is formulated in an optimization framework as finding the parameters that minimize an objective function $f$ on observed data: 
\begin{equation}
    \theta^* = \mathop{\arg\!\min}\limits_{\theta \in \Theta} f(\mathcal{O}, \tilde{\mathcal{O}}_\theta) \text{ or }
    \theta^* = \mathop{\arg\!\min}\limits_{\theta \in \Theta} f(X, \tilde{X}_\theta),
\end{equation}

where the former is most commonly used for calibration of OD matrices while the latter is more commonly used for calibration of CFMs or LC models. That is, CFM and LC models tend to leverage high-fidelity trajectory data (such as the highD or NGSIM datasets) while OD calibration tends to utilize aggregate data (such as detector counts or floating car data) since comprehensive trajectory data at the necessary scales are less common. Examples of prior works that solve the calibration problem formulated as such are detailed in Table~\ref{tab:related_benchmarks}.

\subsection{Calibration metrics}
\label{subsec:calibration_metrics}

Because traffic is stochastic, the goal of calibration is not to match the original data perfectly, but rather to create a microsimulation that reproduces phenomena of interest. This motivates the selection of a particular metric, which should be representative (in some sense) of the behavior the calibration is intended to produce.

The range of possible metrics and the applications they unlock is vast. In their review, \cite{mkadziel2023vehicle} list 17 works that use microsimulation for vehicle emissions applications. \cite{sekar2023micro} review 18 works that leverage simulations for insights into the operational and safety benefits of autonomous vehicles. \cite{raju2021evolution} record 42 papers that leverage microsimulators for modeling connected and automated vehicles.

It is thus infeasible to provide a comprehensive account of these in the present work. However, Table~\ref{tab:calibration_examples} in Appendix~\ref{appendix:microsim_calibs_apps_metrics} provides a non-exhaustive list of microsimulation calibration examples and their associated measures of performance, with a focus on highway settings. This is intended to provide the reader a sense of the possible concrete applications and their diversity. We design our benchmark such that it can accommodate a range of metrics and thus advance research agendas across a broad range of applications.

While conducting this review, we observed the vast majority of these works did not include standard microsimulation best practices described by many works in the following subsection. We believe this further motivates the need for a shared resource between the methods-focused and application-focused communities. In particular, it should be reproducible and rich enough to enable methodological research while also targeting a nontrivial, real-world setting that is useful for applications-focused work. In this way, we envision that methodological enhancements made on our open-source benchmark can be readily and directly adopted by applications-focused researchers and practitioners.

\section{ PROBLEM}
\label{sec:problem}

\subsection{Problem statement}
\label{subsec:problem_statement}

Traffic microsimulation calibration consists of \textbf{selecting} and \textbf{setting} simulation parameters that produce realistic traffic [\cite{hickert2026autotune}]. This work addresses the challenge of establishing a benchmark by which to rigorously and reproducibly compare calibration methods for highway traffic microsimulation.

Because aggregated data (e.g., from induction loops, radar detectors, or floating car data) is significantly more common than more granular, trajectory-level data, we design our benchmark to assess calibration methods that select and set simulation parameters based only on aggregated datasets, as well as those that use trajectory data. The few datasets of high penetration-rate trajectory data that exist (High-D, NGSIM) are limited in temporal and geographic scope. And while INCEPTION v1.0 provides significantly more data -- and will continue to do so in future releases -- this resource remains the exception, rather than the rule. GPS data collected via onboard sensors or smartphone apps are increasingly common, but such data is proprietary and thus largely unavailable to researchers. Furthermore, GPS data does not include every vehicle on the roadway and includes user selection bias. Comparing aggregated-data-only calibration methods to methods utilizing trajectory-level data also will help researchers evaluate the value of collecting such trajectory data, which can be a costly process. 

Since real-world traffic dynamics encompass both vehicle-specific behaviors and emergent macroscopic patterns, this framework is designed to assess simulation fidelity across both microscopic and macroscopic dimensions. Furthermore, trajectory data includes a richer set of features by which to assess traffic reconstructions.

\subsection{Problem Formulation}

For the reader's reference, we provide a notation dictionary in Table~\ref{tab:notation_dictionary}.

\begin{table}[t]
    \centering
    \begin{tabular}{ll}
    \hline
    \textbf{Symbol} & \textbf{Description} \\
    \hline
    $\Theta$ & Parameter space \\
    $\theta$ & Parameters \\
    $c$ & Calibration method \\
    $X$ & Trajectory-level data \\
    $\mathcal{O}$ & Aggregate data \\
    $f$ & Objective function \\
    $\theta^*$ & Optimal parameters \\
    $\tilde{\mathcal{O}}_\theta$ & Simulated aggregate data \\
    $\tilde{X}_\theta$ & Simulated trajectory data \\
    $g$ & Spatial segment \\
    $G$ & Road network \\
    $d_j$ & Distance measures \\
    $\alpha_i, \ldots, \alpha_n$ & Weight terms \\
    $n$ & Number of weight terms \\
    $W_p$ & $p$-Wasserstein distance \\
    $\hat{\mu}^{i,g}_{x_t}$ & Empirical distribution of feature $i$ from $x_t$ \\
    $\hat{\mu}^{i,g}_{\tilde{x}_t}$ & Empirical distribution of feature $i$ from $\tilde{x}_t$ \\
    $t$ & Time \\
    $T$ & Total time evaluations \\
    $\mathcal{C}$ & Set of calibration methods \\
    $c_i$ & $i$-th calibration method \\
    $m$ & Number of calibration methods \\
    $f_i(\cdot)$ & Objective function score \\
    \hline
    \end{tabular}
    \caption{Notation dictionary}
    \label{tab:notation_dictionary}
\end{table}

\subsubsection{Highway Microsimulation Calibration}

We define the \textbf{Highway Microsimulation Calibration} task as the search for the optimal parameter set $\theta^*$ that satisfies the following objective:

\begin{equation}
    \theta^* = \mathop{\arg\!\min}\limits_{\theta \in \Theta} f(X, \tilde{X}_\theta, \mathcal{O}, \tilde{\mathcal{O}}_\theta).
\end{equation}

In this formulation, $f(\cdot)$ serves as a composite loss function aggregating both macroscopic discrepancies and trajectory-level deviations:

\begin{equation}
    f(X, \tilde{X}_\theta, \mathcal{O}, \tilde{\mathcal{O}}_\theta) = \sum_{t \in T} \sum_{g \in G} \sum_{i=0}^m \alpha_i W_p(\hat{\mu}^{i,g}_{x_t}, \hat{\mu}^{i,g}_{\tilde{x}_t}) + \sum_{j=m+1}^n \alpha_j d_j(\mathcal{O}, \tilde{\mathcal{O}}_\theta).
\label{eq:obj_fn}
\end{equation}

In this equation, $G$ represents the complete road network, which is subdivided into discrete spatial regions $g$. To quantify aggregate macroscopic error (e.g., discrepancies in vehicle counts or average velocities), we employ distance metrics $d_j$. The relative importance of each error term is controlled by a set of predefined, normalized weighting coefficients, denoted as $\{\alpha_i, \ldots, \alpha_n\}$, where $\sum_{0}^n \alpha_i = 1$.

To measure microscopic accuracy, we use the $p$-Wasserstein distance, $W_p(\hat{\mu}^{i,g}_{x_t}, \hat{\mu}^{i,g}_{\tilde{x}_t})$, which evaluates the distance between the true empirical distribution ($\hat{\mu}^{i,g}_{x_t}$) and the simulated distribution ($\hat{\mu}^{i,g}_{\tilde{x}_t}$) for a given feature $i$ within segment $g$ at timestep $t$. These features may capture granular state variables like individual inter-vehicle headways, vehicle speeds, or spatial coordinates. Finally, because $\Theta$ is high-dimensional, the resulting loss landscape for $f(X, \tilde{X}_\theta, \mathcal{O}, \tilde{\mathcal{O}}_\theta)$ is highly non-convex and structurally complex.

Dividing the error calculation over network segments $g \in G$ allows us to specify the spatial granularity of the distribution matching -- for example, whether velocity profiles are desired to match across the entire network or at a per-mile basis. Also observe that, to ease computational tractability in practice, objective function evaluation may occur at integer multiples of $t$, such that instead of the outermost summation including $T$ evaluations, it could only require some fraction of $T$ evaluations. 

Importantly, in accordance with~\cite{ciuffo2013no}, this mathematical framing generalizes many typical simulation objectives found throughout the field by subsuming an array of both macroscopic and trajectory-based evaluation metrics and goodness-of-fit criteria [\cite{punzo2021calibration, montali2024waymo}]. To illustrate, researchers can recover standard evaluation setups -- such as calculating a speed distribution error or the root mean square error (RMSE) for sensor counts -- simply by setting specific $\alpha_i$ or $\alpha_j$ terms to 0 and specifying the desired distance metrics. In a similar manner, this architecture also accommodates multi-objective calibration.

A key difference between this formulation and the traditional microsimulation calibrations described in Section~\ref{subsec:traffic_microsim_cal} is that we allow both microscopic and macroscopic data in our objective function. We do this for several reasons. First, as previously described, we seek to take advantage of both forms of data. Aggregate data is relatively plentiful, increasing our benchmark's utility, while trajectory-level data is commonly used in CFM and LC model calibration and is helpful in more granular simulation quality assessment. Furthermore, the I-24 MOTION INCEPTION dataset provides a source of trajectory data that now enables trajectory-level calibration and evaluation at a larger scale than was previously possible. 

Finally, unlike many previous calibration evaluation efforts, we are interested in the full suite of microsimulation parameters and their associated variety of calibration methods. Since some data sources have historically been associated with the calibration of particular parameter sets (as described in Section~\ref{subsec:traffic_microsim}), designing a benchmark with a range of sources is more informative and facilitates better comparison with a wider spectrum of transportation science literature. This includes not only the ability to evaluate the value of the dataset as described above, but also to better understand the relative importance of the simulation parameters and the relationships between them. To the best of the authors' knowledge, we are the first to develop a benchmark that enables such analysis.

This desire to evaluate a range of calibration methods across parameter categories motivates our specific selection of highway traffic on this section of I-24 as the reference, relative to other settings such as urban traffic in the city. In larger, complex networks with free-flowing traffic, we expect overall simulation quality to be driven by the quality of OD parameter calibration and high-fidelity evaluation of the resulting trajectories is difficult. Conversely, on shorter stretches of congested roadway, we expect calibration of CFM and LC model parameters to drive overall performance since the network may be so simple as to have a single origin and destination. Furthermore, while these settings facilitate high-fidelity, trajectory-level evaluation, data availability constraints hamper long-term evaluation. 

Of course, given the scale of real-world highway networks, one may question the value of conducting analysis at the scale of a single-digit mile stretch of roadway. Fortunately, at short to moderate time horizons, highway networks may be approximated by shorter highway segments. That is, highways `look' to drivers like straight lines for the majority of their journey -- there are not the complicated routing considerations and impacts that occur at similar time scales in urban settings. Thus, we hypothesize that methods that perform well on our benchmark may be applicable to larger simulations, assuming the time scale is comparable. 

It is known that traffic behavior is under-specified by aggregated statistics, and thus using only aggregate data as a basis for benchmarking may be an inaccurate measure of the ability of a simulation to reconstruct historical traffic [\cite{wang2024calibrate}]. Given the availability of $X$ on our benchmark, we can leverage this more granular information source as a tool to assist the transportation community as it strives to both develop longer-term and higher-fidelity simulations and also to understand the limitation of those simulations.

\subsubsection{Benchmarking}

Because of the complex topology of $f(\cdot)$, finding the optimal parametrization may be difficult without exhaustive search, which is intractable given the large dimensionality and continuous elements in $\Theta$. Calibration methods have varied approaches to more selectively navigating this search and thus may exhibit different performance.

Given the non-convexity of $f(\cdot)$ mentioned above, locating the global optimum $\theta^*$ analytically is generally impossible, and brute-force searches are intractable due to the continuous and expansive nature of $\Theta$. As a result, researchers deploy various heuristic or data-driven optimization algorithms to traverse this search space. 

Suppose we have a collection of $m$ such highway traffic calibration algorithms, defined as $\mathcal{C} = \{c_1, c_2, \dots, c_m\}$. We frame the task of \textbf{Benchmarking Highway Microsimulation Calibration} as comparing these algorithms by their respective objective scores $f_i(\cdot)$ to rank their effectiveness (where minimizing the score is the goal) [\cite{hickert2026autotune}].

Two challenges arise for assessing calibration methods. First, while multiple trajectories are plausible for a driver, each vehicle in the data only follows one trajectory [\cite{montali2024waymo}]. Put another way, the `true' model for traffic may be distributional, and any historical data represents one draw from that underlying distribution. 
Additionally (and relatedly), the benchmark seeks to maintain construct validity -- namely, that what it measures is of real interest [\cite{reuel2024betterbench}]. 

We can thus analyze, compare, and rank the calibration methods in $\mathcal{C}$ according to their objective function scores $f_i(\cdot)$. This ranking, based on the performance on the test set, provides a means of assessing and comparing the generalization ability of different calibration methods.

\section{ METHODOLOGY}
\label{sec:methodology}

\subsection{Why Highways?}

\begin{figure}[t]
    \centering 
    \begin{subfigure}[b]{0.35\textwidth}
        \centering
        \includegraphics[trim=0mm 0mm 0mm 15mm, clip, width=\textwidth]{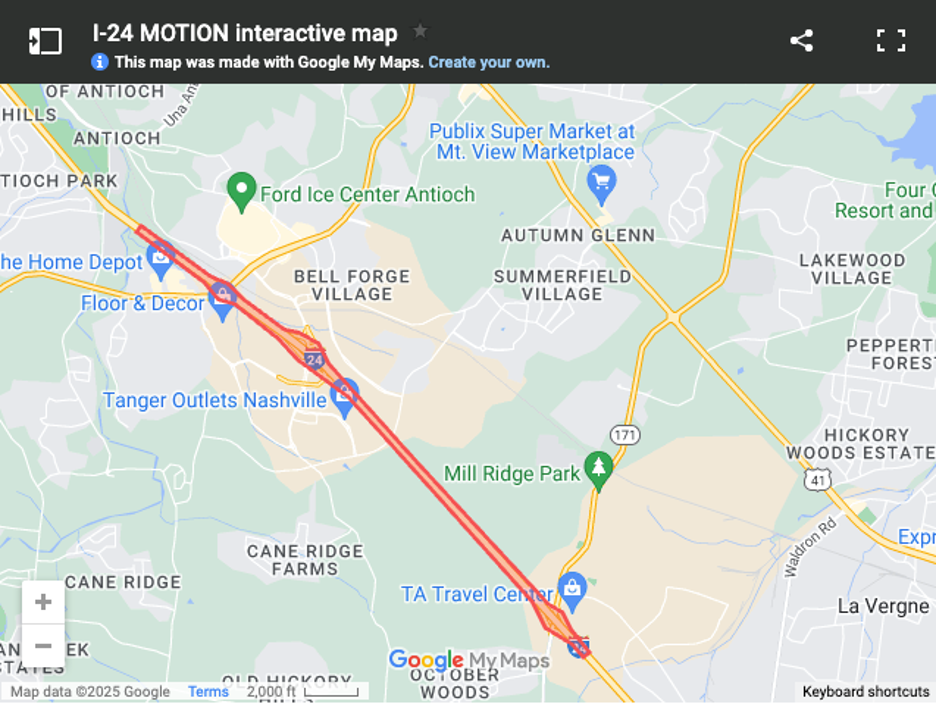}
        \caption{Google Maps view (rights reserved)}
        \label{fig:i24_gmaps_topview}
    \end{subfigure}%
    \hfill
    \begin{subfigure}[b]{0.25\textwidth}
        \centering
        \includegraphics[width=\textwidth]{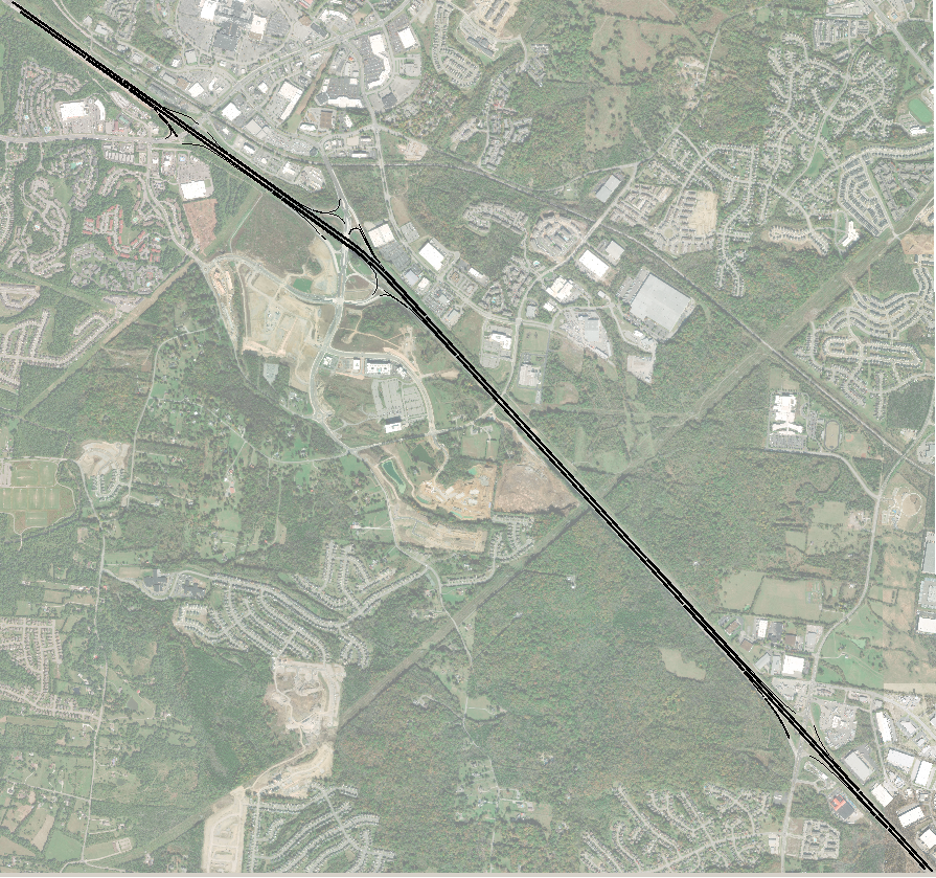}
        \caption{Simulated version (with faded background)}
        \label{fig:i24_sumo_topview}
    \end{subfigure}%
    \hfill
    \begin{subfigure}[b]{0.3\textwidth}
        \centering
        \includegraphics[width=\textwidth]{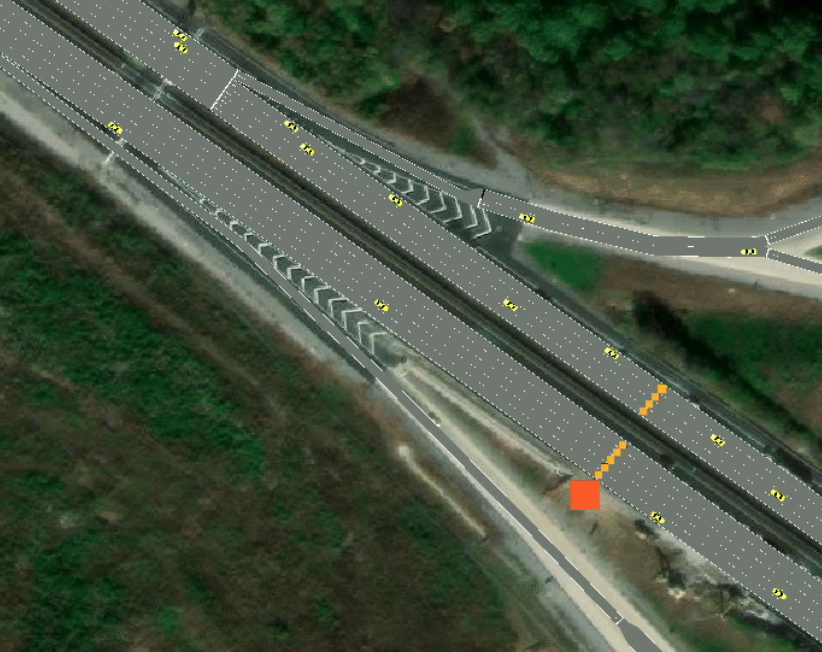}
        \caption{Closer view of simulated network}
        \label{fig:i24_close}
    \end{subfigure}
    \caption{Views of I-24 MOTION in Google Maps and in the Simulation of Urban Mobility (SUMO) traffic microsimulator. Vehicles in the third panel are shown in yellow and `ghost detector' locations are shown in gold with the true detector's physical position shown in orange. Adapted from [\cite{hickert2026autotune}] (© 2026 IEEE).}
    \label{fig:i24_in_sumo}
\end{figure}

Highways offer an ideal testbed for traffic microsimulation for several reasons. First, national and international design standards ensure that highways exhibit similarities consistent across regions [\cite{elefteriadou2024highway}, \cite{trb2000highway}]. A highway calibration framework thus possesses broad applicability to the research community. Furthermore, interstates are important, accounting for 26\% of all vehicle miles traveled in the United States and 15\% of the nation's traffic fatalities [\cite{welki2007impact, national2021early}]. Additionally, their relatively linear topological structure significantly reduces the complexity of origin-destination mapping and vehicle routing. Highways also provide an abundance of macroscopic data from radar, magnetic, and inductive loop sensors. California's PeMS network alone, for instance, utilizes nearly 40,000 detection points to continuously record traffic speeds and volumes at 30-second intervals [\cite{Caltrans2023}]. Finally, the recent release of the I-24 MOTION INCEPTIONv1.0.0 dataset offers 40 hours of high-resolution trajectory data across a 4-mile interstate corridor, yielding a rigorous ground-truth standard for algorithm assessment [\cite{gloudemans202324}] (see Figure~\ref{fig:i24_in_sumo}). As this database expands, it will improve AutoTune's data quantity and quality. By supporting calibration methods trained on both granular trajectory (microscopic) information and traditional aggregate (macroscopic) data, this framework allows practitioners to assess algorithmic performance under realistic deployment conditions and quantify the utility of different data collection modalities.

\subsection{Benchmark Desiderata}
\label{subsec:benchmarking}

\begin{figure}[t]
    \centering \includegraphics[width=0.95\textwidth]{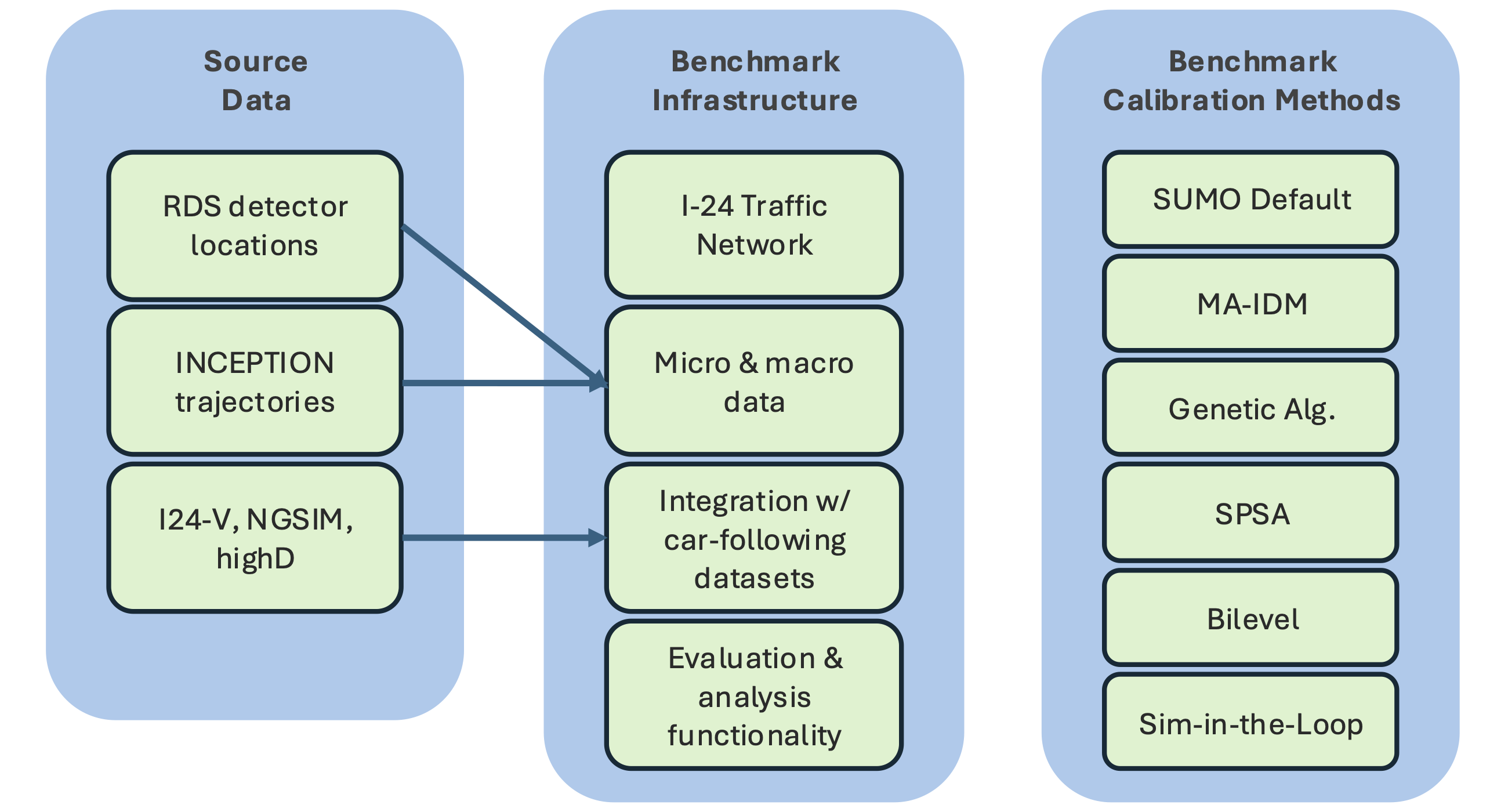} 
    \caption{Conceptual diagram of the benchmark and associated elements. Adapted from [\cite{hickert2026autotune}] (© 2026 IEEE).}
    \label{fig:bench_diagram} 
\end{figure}

We adopt the definition of \cite{raji2021ai} that a benchmark is one or more datasets and metrics representing specific task(s) as a shared framework for a research community to use in comparing methods. While this describes what a benchmark is, it does not describe how best to implement one. For that, AutoTune uses the BetterBench developer guidelines [\cite{reuel2024betterbench}]. Although crafted for artificial intelligence assessments, the BetterBench checklist aggregates best practices from fields as diverse as hardware engineering and bioinformatics. Leveraging the surge in AI evaluation standards provides a strong foundational blueprint for our work, though we substitute the term `scenario' for `task' to better align with the transportation literature. Overall, our architecture addresses 47 best practices spanning the design, coding, documentation, and long-term maintenance phases of the evaluation life cycle. See Appendix~\ref{appendix:benchmark_checklist} for the full checklist and description of AutoTune's performance on each element. Because eight of these criteria are strictly relevant to AI models and not microsimulation calibration, they were excluded. AutoTune satisfies 38 of the 39 that are applicable. The single unmet criterion is the establishment of a human performance baseline; given that manual microsimulation tuning often demands hundreds of labor hours per scenario, AutoTune's design focuses on automatic calibration methods. We leave comparison to human baselines as a possibility for future work. 

\cite{ren2024traffic} also highlight the issue of benchmark saturation, which occurs when methods reach the metric ceiling. Section~\ref{sec:results} demonstrates that AutoTune's large-scale scenarios remain significantly beyond the limits of current algorithmic capabilities. Beyond general benchmarking principles, AutoTune adheres to domain-specific best practices for traffic modeling. This includes utilizing an open-source simulation platform, explicitly defining parameter bounds, and supplying all requisite configuration files. These and others are described in Sections~\ref{sec:related_work} and~\ref{sec:methodology}. An overview of the benchmark and associated elements can be seen in Figure~\ref{fig:bench_diagram}.

\subsection{Benchmark \& Availability}

AutoTune's infrastructure comprises four primary components: (i) simulated SUMO environments mapping to three distinct traffic networks, (ii) data and processing scripts that translate raw I-24 INCEPTION trajectories into macro- and microscopic data formats suitable for calibration, (iii) a standardized evaluation suite, and (iv) supplementary analytical and visualization tools. With permission of the authors, the small and medium networks are modified from \cite{wang2024calibrate} and the large network is modified from \cite{samaei2024integrating}. All code involved in the benchmark will be made available under the MIT license to aid in reproducibility. This includes the SUMO simulation files, data processing modules, calibration methods, error and analysis code, and associated utilities. 

Note that SUMO allows for specification of both simulation step length and reaction time. By default, reaction time is set to step length. But that may or may not be desirable. So for impartiality the way we handle this is to allow each calibration method to select their desired settings for each of these (and/or whether to calibrate one or both of them).

\subsection{Networks}

We evaluated the performance of each method's ability to recreate stop-and-go waves across three scenarios: small, medium, and large. The networks for the small and medium scenarios, respectively, are a 1.3-kilometer on-ramp toy network and a one-way 3-mile segment from the I-24 MOTION network adapted from \cite{wang2024calibrate}. Such toy networks are often used in the literature for assessing calibration methods, while the medium and large networks represent more realistic scenarios. The small network also supports faster iteration for future calibration method development. The large scenario network is a two-way lane-level recreation of the entire 4-mile I-24 MOTION roadway. For the `ground-truth' observed data in each scenario, we used: 8 minutes synthetic aggregate detections (small), 3 hours of radar detection system (RDS) aggregate data (medium), and 12 hours of aggregate data extracted from INCEPTION trajectories (large).

\subsection{Data}
\label{subsec:data}

The benchmark integrates data from three distinct streams. Primarily, it features modules to interact with the INCEPTIONv1.0.0 trajectory dataset [\cite{gloudemans202324}]. Furthermore, we supply both the underlying processing algorithms to generate synthetic loop-detector observations for our designated test scenarios, as well as the extracted data. Finally, we provide interoperability with three different datasets for car-following model calibration, including smoothing to mitigate issues known to occur in computer vision-based trajectory datasets [\cite{coifman2017critical}, \cite{montanino2015trajectory}]. Two of these are well-known (highD and NGSIM); we also integrate with I-24V data, which is collected on the same stretch of freeway as our large network simulates [\cite{gloudemans2024so}].

\subsection{Benchmarked Methods}
\label{subsec:benchmarked_methods}

The benchmark currently evaluates five methods, described below. More details on each method's specific calibration parameters, descriptions, ranges, and initialization values may be found in Appendix~\ref{appendix:cal_params_ranges}.

\begin{figure}[t]
    \centering \includegraphics[width=0.95\textwidth]{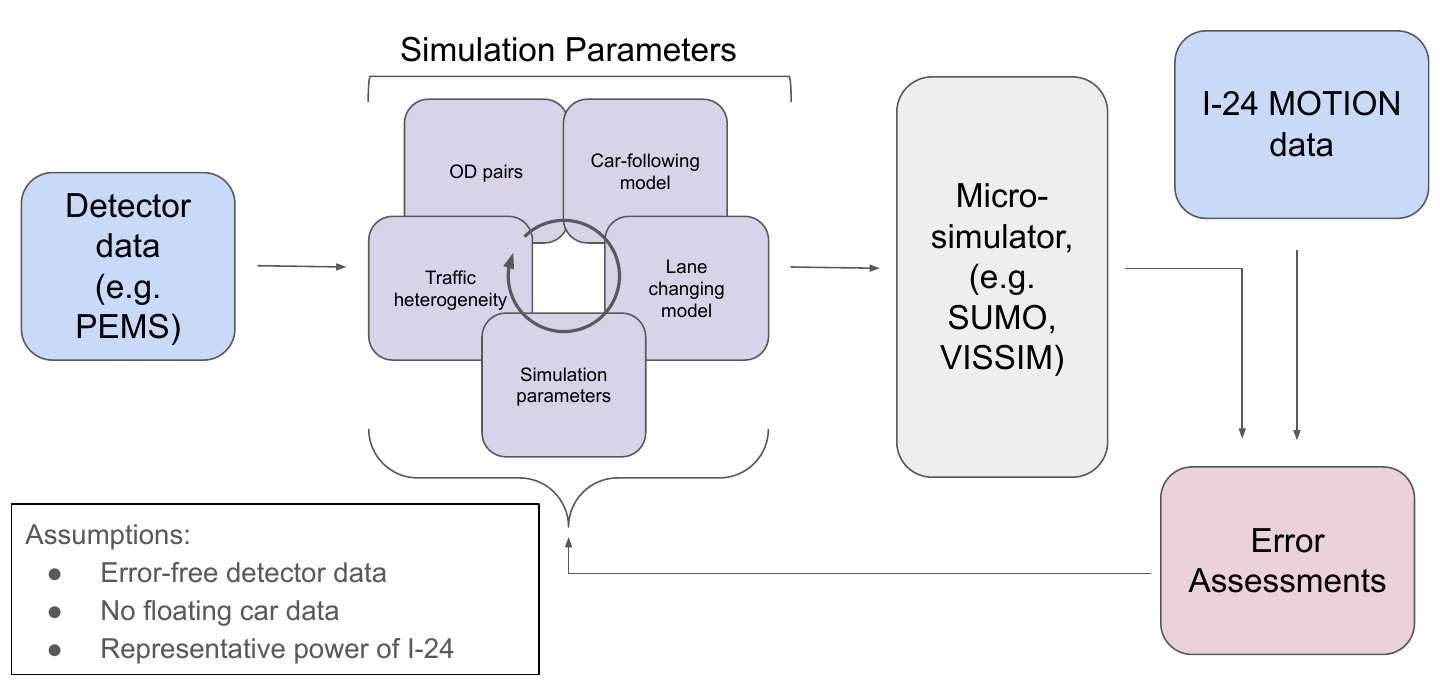} 
    \caption{Overview of microsimulation calibration process.}
    \label{fig:calibration_flow} 
\end{figure}

\subsubsection{SUMO Baseline}

We first provide a baseline using SUMO default parameters where possible. This not only provides an uncalibrated baseline. Given the range of traffic simulation users who report results with minimal calibration, it also helps capture how much error might be incurred in analyses derived from these simulations. This includes the use of the SUMO default car-following model, lane-changing model, and simulation parameters. Heterogeneity is not the default, so vehicles in this method are homogeneous. SUMO defaults for simulation parameters are a one-second simulation timestep length, no warm-up time, and a continuous simulation (rather than stitching together multiple shorter simulations). 

The one calibrated component of this baseline is the OD matrix. This is done using SUMO's built-in `Flowrouter' functionality, which is designed to construct OD flows from incomplete detector data -- that is, networks without full detector coverage, such as the one constructed in this work. Flowrouter infers traffic on edges from nearby detection data. Doing so requires the user to specify an aggregation interval; we set it to be an hour. 

\subsubsection{Memory-Augmented IDM (Bayesian Calibration)}
Bayesian calibration is included in this benchmark for its theoretical rigor, built-in uncertainty quantification, and increasing relevance in safety-critical and interpretable traffic simulation applications [\cite{zhang2024bayesian, zhang2024calibrating}]. Traditional calibration methods, such as Genetic Algorithms (GA) or other heuristic optimizers, often yield only a single point estimate of the model parameters by minimizing a loss function. While these point estimates can reproduce average traffic patterns, they ignore parameter uncertainty and may lead to overconfident or unstable simulation outcomes, especially when deployed in previously unseen scenarios or for downstream tasks like autonomous vehicle testing.

By contrast, the Bayesian framework provides a distribution over plausible parameters given observed data and a probabilistic model of the traffic dynamics. This not only captures epistemic uncertainty in the model parameters through the posterior distribution, but also aleatoric uncertainty through a stochastic error model that accounts for temporally correlated residual variability in driving behavior. Together, these allow for robust and uncertainty-aware decision-making, particularly important in safety-critical traffic simulation and policy testing. The resulting posterior distributions allow analysts to identify which parameters are well-constrained by data, assess identifiability, and propagate uncertainty through to simulation outputs.

This method specifically targets the calibration of Intelligent Driver Model (IDM) parameters using trajectory-level data. A likelihood function is defined based on the discrepancy between observed and simulated trajectories (e.g., in terms of spacing, speed, and acceleration). Importantly, this method extends beyond standard independent-error assumptions by incorporating a stochastic error model (e.g., Gaussian processes [\cite{zhang2024bayesian}] or autoregressive processes [\cite{zhang2024calibrating}]) that captures temporally correlated residuals. Such correlations arise naturally in traffic flow: driver actions are not independent from one moment to the next, and sensor or modeling errors accumulate over time. Accounting for these correlations leads to better-fitting models, more realistic trajectory reconstructions, and tighter uncertainty bounds.

Technically, the method is implemented via Markov Chain Monte Carlo (MCMC) sampling, which approximates the joint posterior distribution over both IDM parameters and hyperparameters of the residual process (e.g., Gaussian process lengthscale and variance). This enables the method to accommodate both fixed parameters (e.g., the acceleration exponent in IDM) and temporally correlated noise in the residual dynamics, thereby modeling persistent patterns in driver behavior that unfold over time.

The approach is fully probabilistic and yields not only point estimates but full posterior distributions and predictive intervals, enhancing its suitability for applications that demand interpretability and reliability. These include risk-aware autonomous driving simulations, safety validation under uncertainty, transferability testing across traffic regimes, and policy evaluation on high-fidelity datasets like I-24 Motion [\cite{gloudemans2024so}].

This Bayesian approach calibrates key IDM parameters such as desired speed, time headway, minimum spacing, maximum acceleration, and comfortable deceleration. Additionally, the residual error process is modeled with a Gaussian process, whose lengthscale and variance hyperparameters are also inferred. These are detailed in Table~\ref{tab:MA-IDM-params}.

\subsubsection{Genetic Algorithm (GA)}

As shown in Table~\ref{tab:related_benchmarks}, GA is the most popular calibration methodology across categories. A basic implementation of a genetic algorithm is included in this benchmark due to it's widespread use in traffic micro-simulation [\cite{ciuffo2013no}]. Genetic algorithms are also relatively simple to implement while offering good results. This method uses a genetic algorithm to calibrate the car following parameters in the IDM model, the lane changing parameters, the simulation parameters, and the heterogeneity parameters. Due to the ubiquitous nature of genetic algorithms, a researcher that wants to try out their method but lacks calibration of certain parameters can simply use a genetic algorithm as a base calibration for parameters not considered by their method.

For both GA and SPSA, we adopt a block optimization approach as seen in Figure~\ref{fig:calibration_flow}. This supports the joint consideration of parameters described previously. 

\subsubsection{SPSA}
Simultaneous Perturbation and Stochastic Approximation (SPSA) was also used as a base model. SPSA estimates the gradient of a non-differentiable function by randomly perturbing an input in random directions, evaluating a cost function, and moving the solution in the direction of lower cost. It is frequently used in traffic microsimulation [\cite{ciuffo2013no}], because traffic microsimulations are noisy and stochastic, and SPSA is a computationally efficient way to minimize an objective function.

\subsubsection{Simulation-in-the-loop optimization}
This is a straightforward simulation-in-the-loop calibration framework designed to identify optimal combinations of microscopic driving behavior parameters, including car-following and lane-changing, regardless of the specific model forms or number of parameters involved. The framework adopts a one-step approach that relies solely on easily accessible macroscopic data from stationary sensors, such as flow, speed, and occupancy. Calibration is formulated as an optimization problem that seeks to minimize discrepancies between simulated and observed traffic states over a defined spatiotemporal domain. A global optimization algorithm, Differential Evolution, is employed to navigate the non-convex, simulation-based objective landscape. As demonstrated in~\cite{wang2024calibrate}, jointly calibrating both car-following and lane-changing parameters using speed or occupancy data most effectively reproduces critical traffic phenomena, including congestion onset, wave propagation, and lane-specific travel time patterns.

A major strength of this framework is its practicality: it leverages widely available macroscopic sensor data and does not require high-resolution trajectory datasets, making it particularly suitable for real-world applications with limited sensing infrastructure. The approach is also flexible—users define the microscopic models and parameter sets externally in SUMO, without needing model-specific or optimization-specific modifications. However, the method is computationally intensive and did not consistently achieve convergence within the preset iteration limits. It also assumes a homogeneous driver population, which may limit its ability to capture realistic behavioral heterogeneity. Furthermore, since the framework is designed to replicate macroscopic traffic features, it does not explicitly ensure accuracy at the individual vehicle level. A predefined, time-varying origin–destination (OD) demand profile must be supplied, which highlights the modular nature of the framework but also the importance of accurate demand estimation for successful calibration.

\section{ EXPERIMENTS}
\label{sec:experiments}

To assess the performance of the algorithms on a range of real-world settings, we define one scenario on the small network, one on the medium network, and three scenarios on the large network. Each is defined by its objective, measure of performance, goodness of fit function, and data scope. Those for the large network are shown in Table~\ref{tab:scenarios} and are chosen because traffic simulations are an important tool for traffic theory, application-focused research, and are widely used in government and industry applications. The speed and count (flow) aggregate error terms are common in highway calibration (see Tables~\ref{tab:related_benchmarks} and~\ref{tab:calibration_examples}. For Scenario 2, the objective function is populated with headway distribution microscopic error terms as recommended in \cite{punzo2021calibration}. The MoPs used also provide useful diversity: the microscopic metric compares distributions of vehicle-level headway values, the intermediate metric compares speeds averaged across uniform `windows' comprised of 10-second durations on 400m directional road segments, and the macroscopic metric only uses data at the ghost detector locations.  Microscopic error function segments are a mile long and the function is evaluated at 15-minute increments. Average Wasserstein distance and MAPE are metrics to be minimized, with target values of $0$. 

Multiple random seeds for each method are run to account for stochasticity in the techniques and the underlying SUMO simulator. Results are reported as means and standard deviations across metrics. While these scenarios are provided as useful assessment tools, we also note that the benchmark supports user definition and data processing for arbitrary scenarios based on the INCEPTION data. 

\begin{table}[t]
    \centering
    \begin{tabular}{|p{2.5cm}|p{3.5cm}|p{3.5cm}|p{3.5cm}|}
    \hline
    \textbf{Use type} & Theoretical research & Application-focused research & Government \& industry applications \\
    \hline
    \textbf{Goal} & Informing traffic theory on congestion propagation & Safety assessment & Throughput assessment \\
    \hline
    \textbf{MoP Type} & Intermediate error metric & Microscopic error metric & Macroscopic error metric \\
    \hline
    \textbf{MoP} & Per-segment (400m, 10s) average speeds & Per-vehicle headway distributions & 30-second average counts (flows) at detectors \\
    \hline
    \textbf{GoF Function} & MAPE & Average Wasserstein Distance & RMSE \\
    \hline
    \textbf{Data Scale} & Small & Medium & Large \\
    \hline
    \textbf{Data Scope} & 7-9AM on Monday, 11/21/2022, and Monday, 11/28/2022. & 6-10AM on Monday, 11/21/2022, Tuesday, 11/22/2022, Tuesday, 11/29/2022, and Friday, 12/02/2022. Crashes occurred on the 21st and 2nd. & 6-10AM on Monday-Friday, 11/21/2022 - 11/25/2022, Monday-Friday, 11/28/2022 - 12/02/2022. The 24th and 25th are holidays (Thanksgiving and day after Thanksgiving). \\
    \hline
    \end{tabular}
    \caption{Overview of scenarios on the large network. Those for the small and medium networks are described in the main body.}
    \label{tab:scenarios}
\end{table}

Performance for these is assessed using the INCEPTIONv1.0.0 data as `ground truth' to enable fair comparisons. To assess the impact of sensor error, we will  later provide Scenario 6, which is the same as Scenario 5 but applies realistic noise to the aggregated detection data. This process is described in detail in Appendix~\ref{appendix:det_noise}.

To improve understanding of how scale affects the difficulty of highway calibration, relative to the simplified highway calibration settings under which methods are commonly developed. We assess performance as a function of simulation size, itself comprised of spatial magnitude, number of vehicles, and temporal duration. 

To do so, we add two scenarios to those described above, using additional small and medium networks, described above and sourced from \cite{wang2024calibrate}. We designate the small network case as Scenario 1 and the medium network case as Scenario 2. Each uses the same MoP and GoF Function as Scenario 3: MAPE on the per-segment average speeds, as is common in congestion analysis. Both scenarios use synthetic data as ground truth, guaranteeing that the data-generating process can be simulated and allowing us to exclude issues that arise due to data collection error. The small network duration is eight minutes and includes 400 vehicles, while the medium network is one hour and includes 4,000 vehicles. 

The inclusion of a synthetic scenario also presents an opportunity to investigate the role of detector error. To do so, we select the best-performing calibration method from the small network calibration and re-calibrate against data with 5\% Gaussian noise applied. 

Furthermore, we parse and extract objective values, best-fitness traces, and block-specific performance for parameter groups such as demand, car-following, lane-changing, simulation, and heterogeneity parameters from GA and SPSA calibration traces. This allows us to compare which parameter blocks drive improvements.

Finally, we use small-scale microsimulation to demonstrate the performance sensitivity of calibration methods to parameters not included in them (e.g., simulation timestep).

\section{ RESULTS}
\label{sec:results}

Table~\ref{tab:exp1_smallnet} summarizes the small network evaluation across SUMO random seeds. For each method, we report mean $\pm$ standard deviation for detector count RMSE, headway distribution error, and velocity-grid MAPE. The GA and SPSA runs were calibrated directly against count RMSE, while MA-IDM and the SUMO baseline are fixed-method comparisons; all methods are then evaluated using the same set of metrics. These results therefore illustrate one purpose of the benchmark and provide a comparative assessment on this setting. A method can improve the objective it is tuned for while performing differently on other traffic-state quantities, making multi-metric evaluation useful for diagnosing calibration tradeoffs rather than relying on a single final score. Additional benchmarking results for the medium and large networks will be updated as the working paper matures. Further refinements and more granular result reporting for the small network will also be included. Generally, we note that methods calibrating more parameter sets (GA, SPSA) outperform those calibrating fewer parameters, indicating the importance of broad parameter inclusion.

\begin{table}[ht]
    \centering
    \caption{Performance across methods and metrics for the selected small network scenario. (Lower is better.)}
    \label{tab:exp1_smallnet}
    \begin{tabular}{lccc}
        \toprule
        Method & Count RMSE & Headway Wass. & Velocity MAPE \\
        \midrule
        SUMO Default & 5.055 $\pm$ 0.074 & 30.665 $\pm$ 4.983 & 39.844 $\pm$ 12.997 \\
        GA & 4.404 $\pm$ 0.700 & 23.449 $\pm$ 6.917 & 9.588 $\pm$ 1.866 \\
        SL & 3.705 $\pm$ 0.658 & \textbf{18.119} $\pm$ \textbf{5.707} & \textbf{5.200} $\pm$ \textbf{1.519} \\
        MA-IDM & 5.933 $\pm$ 0.070 & 27.830 $\pm$ 5.100 & 32.935 $\pm$ 15.830 \\
        SPSA & \textbf{3.590} $\pm$ \textbf{0.406} & 23.498 $\pm$ 5.082 & 18.084 $\pm$ 0.875 \\
        \bottomrule
    \end{tabular}
\end{table}

\begin{figure}[th]
    \centering
    \begin{subfigure}[b]{0.3\textwidth} 
        \centering
        \includegraphics[width=\textwidth]{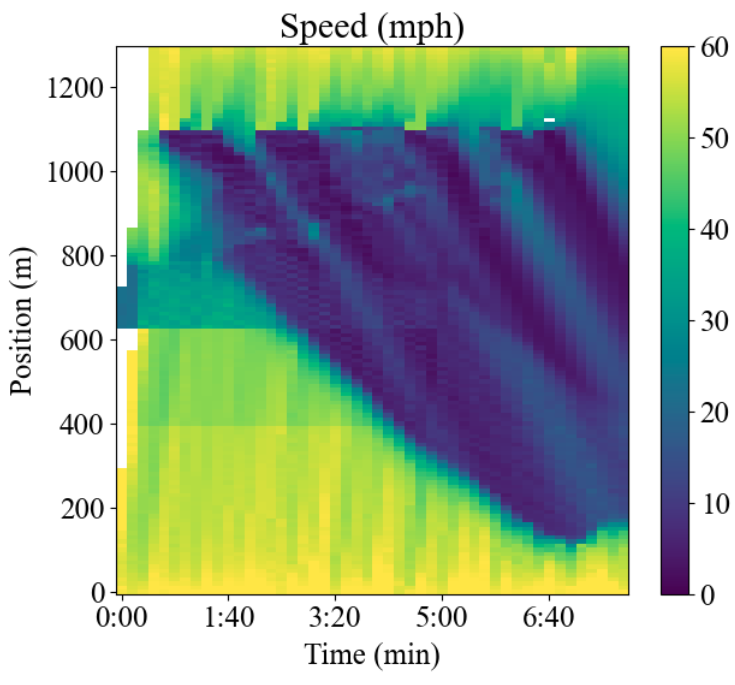}
        \caption{Small scenario data (synth.)}
        \label{fig:speed_smallnet_gt}
    \end{subfigure}
    \hfill
    \begin{subfigure}[b]{0.3\textwidth}
        \centering
        \includegraphics[width=\textwidth]{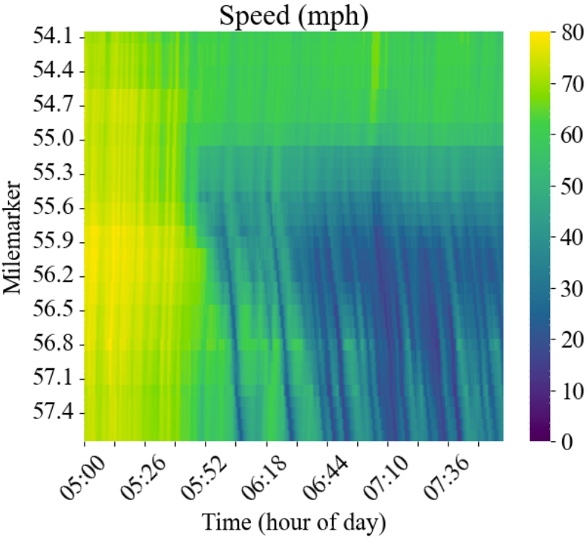} 
        \caption{Medium scenario data}
        \label{fig:speed_mednet_gt}
    \end{subfigure}
    \hfill
    \begin{subfigure}[b]{0.3\textwidth}
        \centering
        \includegraphics[width=\textwidth]{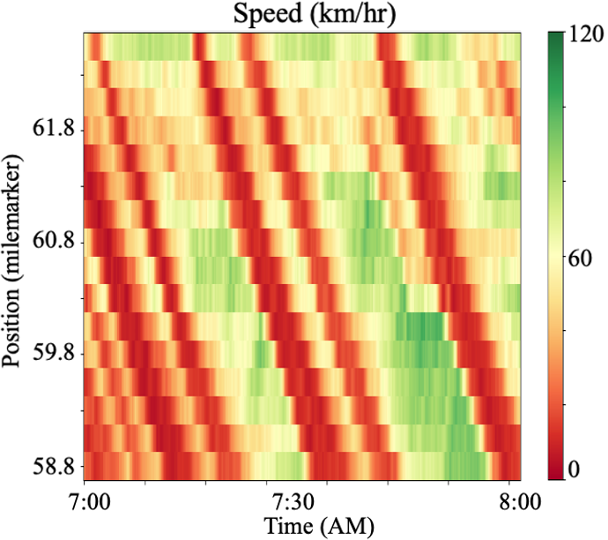}
        \caption{Large scenario data}
        \label{fig:speed_largenet_gt}
    \end{subfigure}
    
    \vspace{0.5cm} 
    
    \begin{subfigure}[b]{0.3\textwidth}
        \centering
        \includegraphics[width=\textwidth]{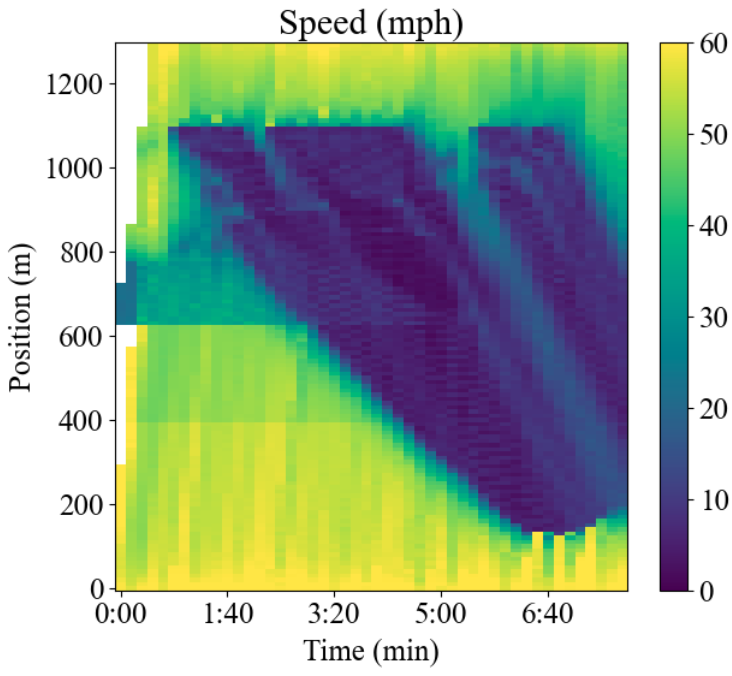} 
        \caption{Small scenario simulation}
        \label{fig:speed_smallnet_sim}
    \end{subfigure}
    \hfill
    \begin{subfigure}[b]{0.3\textwidth}
        \centering
        \includegraphics[width=\textwidth]{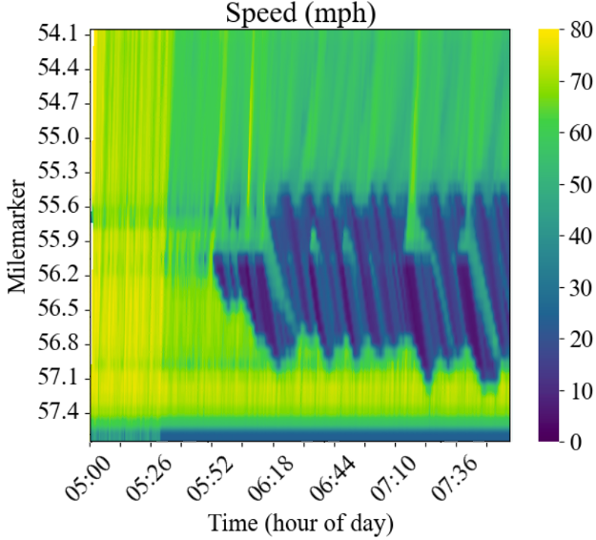}
        \caption{Medium scenario simulation}
        \label{fig:speed_mednet_sim}
    \end{subfigure}
    \hfill
    \begin{subfigure}[b]{0.3\textwidth}
        \centering
        \includegraphics[width=\textwidth]{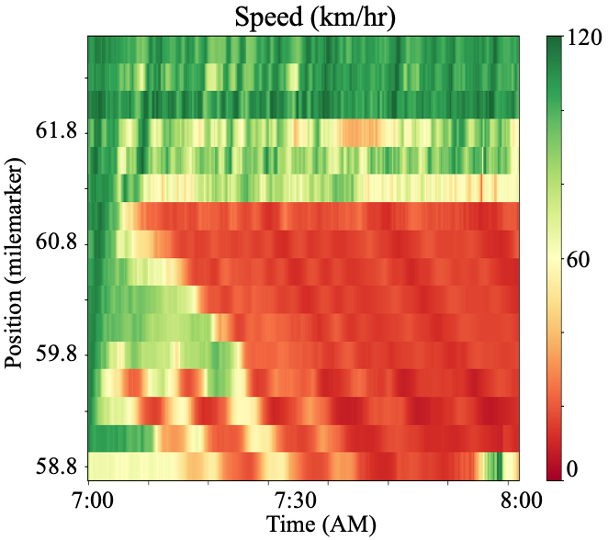} 
        \caption{Large scenario simulation}
        \label{fig:speed_largnet_sim}
    \end{subfigure}
    \caption{The validity of the stop-and-go waves produced in the calibrated microsimulations degrades as scenario complexity grows. Medium scenario results are reproduced from \cite{wang2024calibrate}.}
    \label{fig:stop_and_go_degradation_w_net_size}
\end{figure}

\begin{figure}[h!]
    \centering 
    \begin{subfigure}[b]{0.45\textwidth}
        \centering
        \includegraphics[trim=0cm 0cm 0cm 1cm, clip, width=\textwidth]{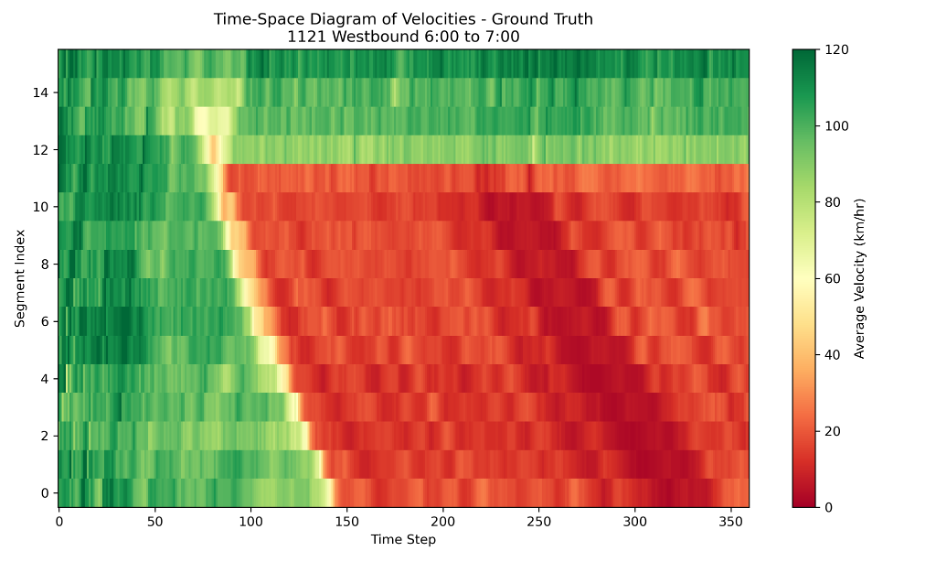}
        \caption{Ground truth}
        \label{fig:ts_congest_day_gt}
    \end{subfigure}%
    \hspace{0cm} 
    \begin{subfigure}[b]{0.45\textwidth}
        \centering
        \includegraphics[trim=0cm 0cm 0cm 1cm, clip, width=\textwidth]{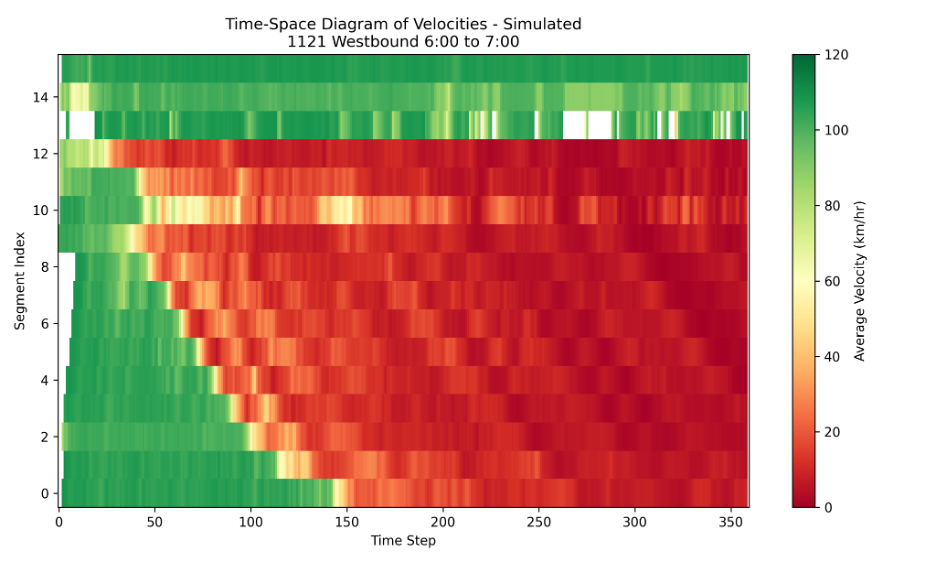}
        \caption{Simulated}
        \label{fig:ts_congest_day_sim}
    \end{subfigure}%
    \caption{Effective speed simulation on a day with consistent congestion.}
    \label{fig:ts_congest_day}
\end{figure}

\begin{figure}[t]
    \centering 
    \begin{subfigure}[b]{0.4\textwidth}
        \centering
        \includegraphics[trim=0cm 0cm 0cm 1cm, clip, width=\textwidth]{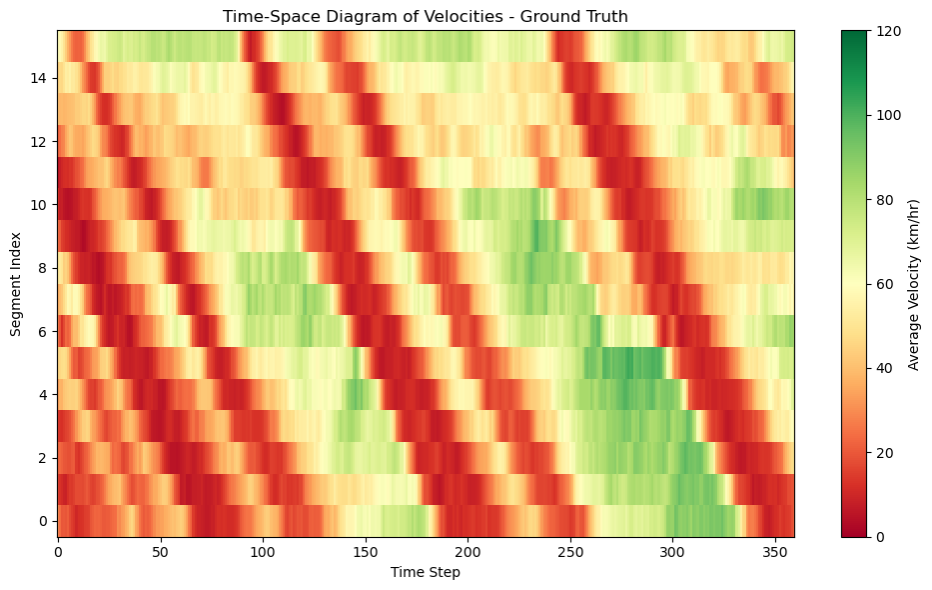}
        \caption{Ground truth}
        \label{fig:ts_stopandgo_day_gt}
    \end{subfigure}%
    \hspace{0cm} 
    \begin{subfigure}[b]{0.4\textwidth}
        \centering
        \includegraphics[trim=0cm 0cm 0cm 1cm, clip, width=\textwidth]{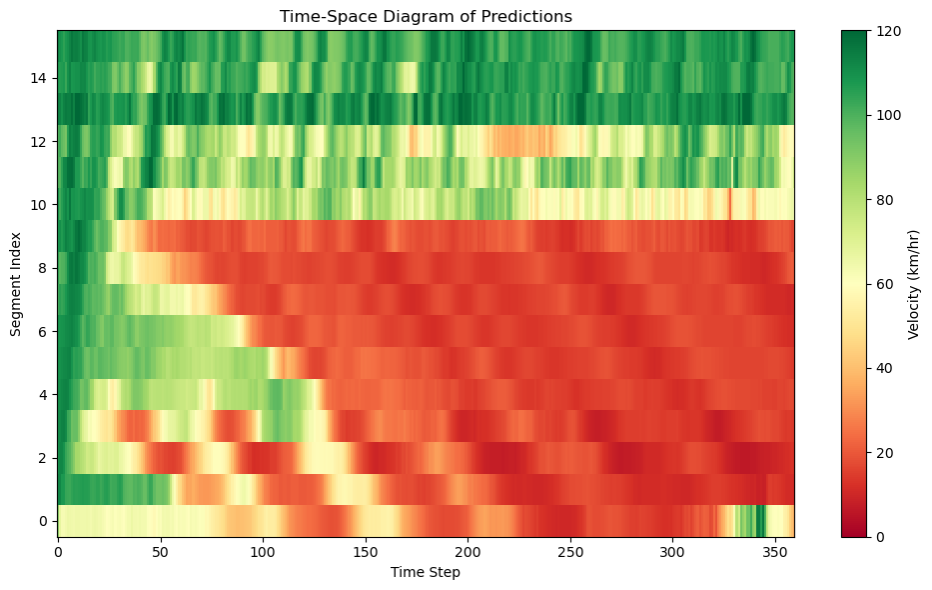}
        \caption{Simulated}
        \label{fig:ts_stopandgo_day_sim}
    \end{subfigure}%
    \caption{Poor performance on a day with stop-and-go waves.}
    \label{fig:ts_stopandgo_day}
\end{figure}

The speed fields resulting from the best of the six methods on each simulation can be seen in Figure~\ref{fig:stop_and_go_degradation_w_net_size}. Only one hour of results for the large scenario are shown for clarity, but the pattern persists. These reveal the performance gap component of the calibration illusion for microsimulation: namely, the validity of the simulations is insufficient for the purpose of recreating stop-and-go waves at large scale. Furthermore, the panel of results from the small to large cases exhibits that performance declines as scenario complexity grows (in terms of network size, number of vehicles, and duration).

This illustrates that, when evaluated transparently on scenarios of real-world interest, significant gaps remain between observed data and the microsimulations produced by current methods. Notably, this performance is lower than the literature implies [\cite{samaei2024integrating, amirjamshidi2019multi}]. This deficiency is particularly detrimental for assessing stop-and-go waves: unlike use cases like throughput assessment or headway distributions where qualitative behaviors can remain even as quantitative errors grow, this manifests as a distinct qualitative failure where waves degrade to \textbf{non-existence}. These results highlight both (i) the gaps that remain for traffic microsimulation calibration on real-world settings of interest relative to the simplified highway calibration settings under which methods are commonly developed, and (ii) the importance of assessing techniques on realistic settings. 

We find high variance in even the best method's ability to recreate qualitative behavior of interest. Figure~\ref{fig:ts_congest_day} depicts a situation in which the genetic algorithm is able to reasonably recreate ground-truth data in a consistently congested setting, while Figure~\ref{fig:ts_stopandgo_day} illustrates a setting where the desired behavior is not replicated. Overall, we find all methods are insufficient to capture stop-and-go behavior of traffic waves. The results are more mixed for the other metrics of interest, perhaps in part because other measures of performance do not have as binary a result as the ability to recreate stop-and-go waves. Given the vast range of possible applications of traffic microsimulation (further described in Table~\ref{tab:calibration_examples}), we simply present results here and allow the research and practitioner community to decide whether the performance is sufficient for their needs. Alongside the open-source implementations and associated documentation, we believe this can greatly ease user adoption and adaptation of the methods and settings.

\begin{figure}[t]
    \centering 
    \begin{subfigure}[b]{0.3\textwidth}
        \centering
        \includegraphics[width=\textwidth]{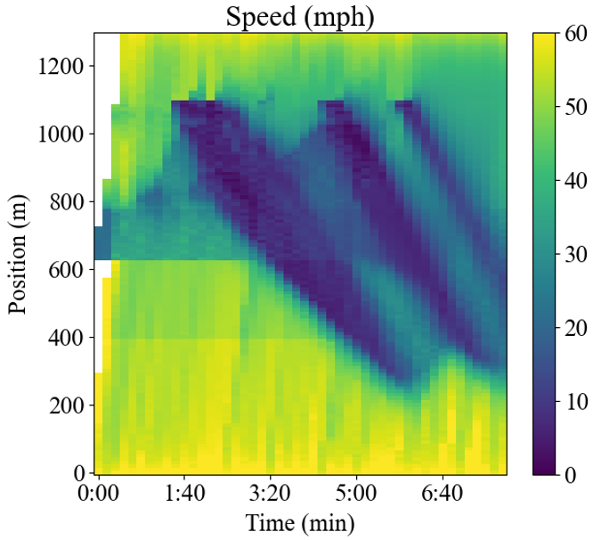}
        \caption{Time step = 0.5 sec.}
        \label{fig:smallnet_01sectstep}
    \end{subfigure}%
    \begin{subfigure}[b]{0.3\textwidth}
        \centering
        \includegraphics[width=\textwidth]{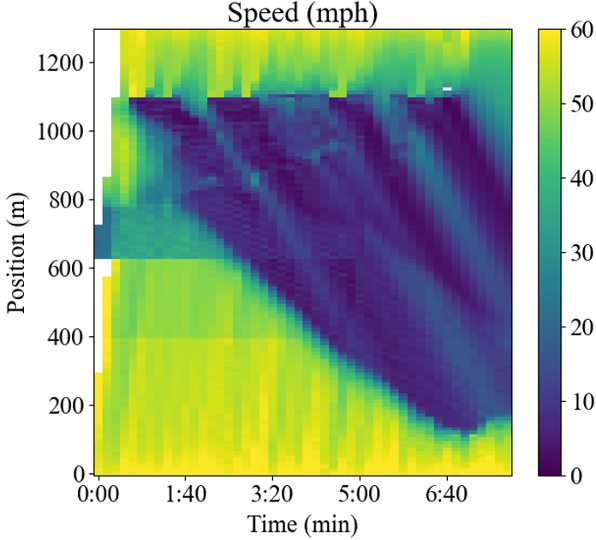}
        \caption{Time step = 0.55 sec.}
        \label{fig:smallnet_02sectstep}
    \end{subfigure}%
    \caption{Parameters traditionally unspecified and uncalibrated can have significant impacts on the recreation of desired phenomena. Shown is the difference in traffic speed wave phenomena due to a 10\% change in the simulation's time step parameter, while all other parameters are fixed.}
    \label{fig:tstep_comparison}
\end{figure}

\begin{figure}[th]
    \centering
    \includegraphics[trim=0cm 0cm 0cm 0cm, clip, width=0.9\linewidth]{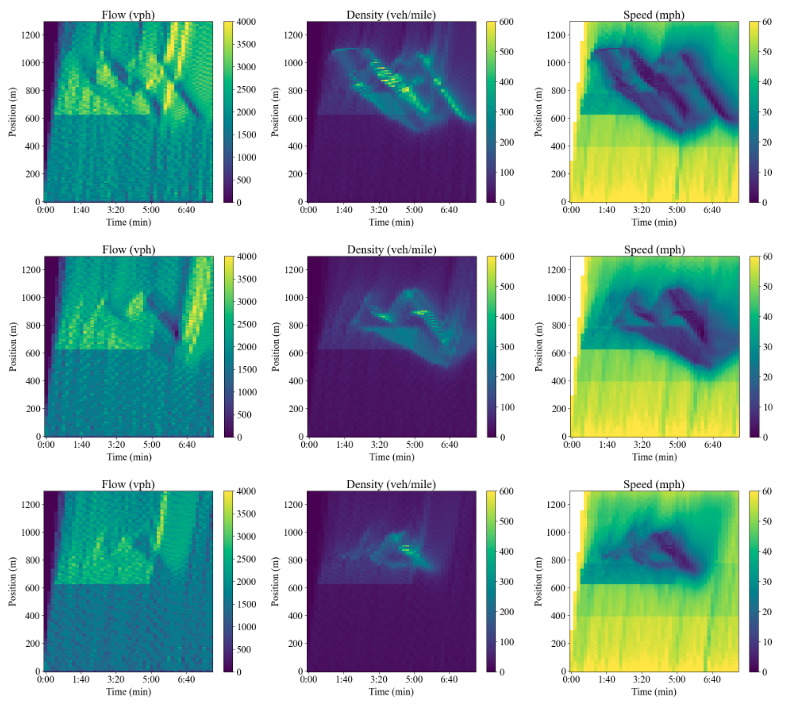}
    \caption{As sensor error (noise) increases, the quality of simulation quickly degrades, even for the small network. Flow, density, and speed plots in the top row are shown for the synthetic ground truth, followed by results from the calibrated simulation for that data in the middle row. The bottom row shows the results when 5\% noise is added to the data used in calibration.}
    \label{fig:small_net_noise_impact}
\end{figure}

\begin{figure}[th]
    \centering
    \includegraphics[width=0.7\linewidth]{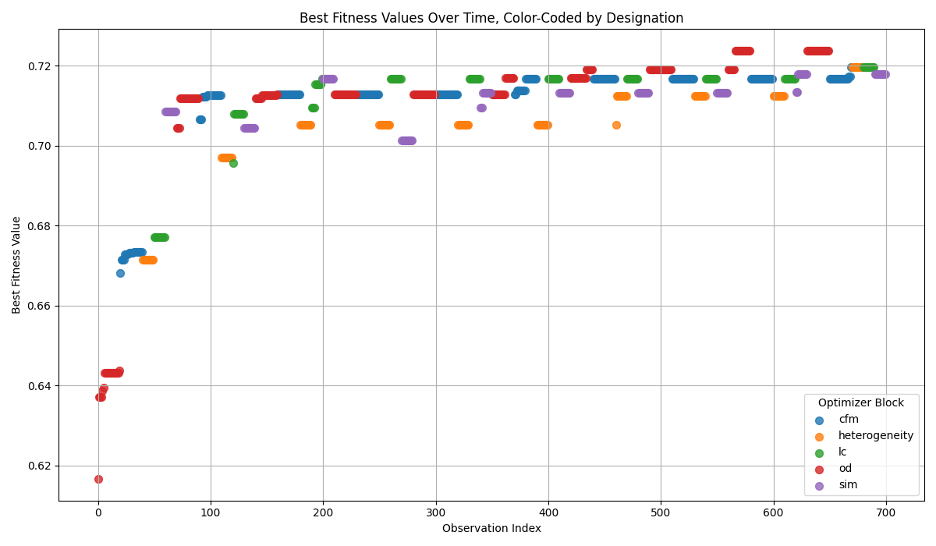}
    \caption{An illustration of how the various parameter categories contribute to the overall growth in performance on the large network.}
    \label{fig:cal_curve_by_pararm_cat}
\end{figure}

We also investigated the manual dependence of each method. We found that minor (10\%) adjustments to parameters traditionally excluded from calibration (such as individual simulation and heterogeneity parameters) could more than \textbf{double} the resulting mean absolute percent error (MAPE) of the simulation's speed fields. See Figure~\ref{fig:tstep_comparison} for one example. This shows that two different researchers using the same ‘method’ can get very different results if that method does not fully specify the simulation, and even potentially abandon promising methods. It is thus important to consider how calibration methods tune or set all parameters that affect performance.

Furthermore, in scenarios studied, the calibration of these traditionally excluded categories contributed significantly to the overall performance gain achieved in the calibration process; this demonstrates that more holistic parameter calibration can improve performance. See Figure \ref{fig:cal_curve_by_pararm_cat}. Overall, this indicates that common calibration methods in the literature are underspecified for highway microsimulation. Their performance is sensitive to parameters that are not described in the methods’ initial presentation or use. This reduces reproducibility and can misrepresent their performance on the task in both absolute terms and relative to the performance of other methods. Including these parameters in the calibration process can improve microsimulation quality. Doing so in a block optimization framework enhances tractability in large-scale simulation scenarios, such as those leveraging data from the I-24 MOTION testbed.

A related factor raising the difficulty of calibration is the presence of error that occurs as real-world detector measurements are used. This is illustrated in Figure~\ref{fig:small_net_noise_impact}. This investigation utilizes the small setting, where synthetic ground truth is available. In contrast, the other settings use real-word data and thus are likely to already include error in their measurements.

\clearpage

\section{ CONCLUSION}
\label{sec:conclusion}

This working paper identifies and characterizes the automatic calibration illusion in traffic microsimulation. This demonstrates that this illusion obscures a validity gap, where supposedly automatic methods rely on undocumented manual tuning and fail to reproduce critical phenomena like stop-and-go waves. To address these distinct failures, we introduce AutoTune to enforce transparency and reduce research siloing in microsimulation. We demonstrate how more holistic parameter calibration can improve performance, although a gap remains for the recreation of stop-and-go waves on the large scenario. This work shifts microsimulation calibration from opaque, manual tuning toward transparent, more robust methodologies.

\appendix

\section{Appendix: In-Depth Comparison of Related Work}
\label{appendix:related_work}

Table~\ref{tab:related_benchmarks} assesses related work along a variety of attributes. Where applicable, these are summarized with indicators for yes (\yes), no (\no), and somewhat(\soso). The characteristics evaluated are:
\begin{itemize}
    \item Benchmark: Does the work score and compare multiple calibration approaches (incl. algorithm, metric, or simulator) to evaluate performance quality on a common setting? This is essential for conducting a fair cross-method evaluation and comparison.
    \item OD/CFM/LC/Simulator parameters: Are parameters in the respective parameter category subject to calibration? This bounds which calibration problems may be considered and informs a work's ability to jointly consider parameters, as described above. 
    \item Heterogeneity: Is non-homogenous traffic modeled? E.g., varied driver types or vehicle lengths. Real-world traffic is heterogeneous. 
    \item Microscopic data: Is data at the level of individual vehicles used in the calibration process? This is common among CFM and LC model calibration methods. It enables higher-fidelity analysis than macroscopic data alone.
    \item Aggregated (macroscopic) data: Is data at the traffic-level (i.e., pooled across multiple vehicles) used in the calibration process? This is common among larger-scale simulation calibration and is helpful for capturing traffic-level phenomena. Aggregated data is also more plentiful than microscopic data, and thus methods which use it have wider applicability. 
    \item Microsimulator: Which microsimulator was used. Note that many CFM calibration methods only require leader-follower trajectory pairs, and thus may not require a microsimulator during the calibration process (even if rigorous assessment of their performance at the traffic-level requires it). 
    \item Setting: The road type included. 
    \item Source data: The type of data used for calibration.
    \item Data scale: The amount of data used for calibration. 
    \item MoP (Measure of performance): The measured quantity being used to compare simulations with the source data. E.g., detector counts over time, inter-vehicle spacing, etc. 
    \item GoF (Goodness-of-fit function): The mathematical expression by which the simulated and real MoPs are compared. E.g., RMSE, MAE, etc. 
\end{itemize}

\begin{landscape} 

    \begin{longtable}{ |p{2cm}|p{2cm}|p{1.5cm}|p{1.5cm}|p{1.5cm}|p{2.1cm}|p{1.5cm}|p{1.5cm}|p{1.6cm}|p{1.2cm}| p{2.5cm}| }
    
    \caption{Related benchmarks, reviews, and studies}
    \label{tab:related_benchmarks} \\
    \toprule
    Paper & \cite{kim2005calibration} & \cite{punzo2021calibration} & \cite{hollander2008principles} & \cite{montali2024waymo} & \cite{amirjamshidi2019multi} & \cite{chen2023follownet} & \cite{punzo2012can} & \cite{ciuffo2013no} & \cite{samaei2024integrating} & Ours \\
    \midrule
    \midrule
    \endfirsthead
    
    \multicolumn{4}{c}%
    {{\bfseries Table \thetable\ continued from previous page}} \\
    \toprule
    Paper & \cite{kim2005calibration} & \cite{punzo2021calibration} & \cite{hollander2008principles} & \cite{montali2024waymo} & \cite{amirjamshidi2019multi} & \cite{chen2023follownet} & \cite{punzo2012can} & \cite{ciuffo2013no} & \cite{samaei2024integrating} & Ours  \\
    \midrule
    \midrule
    \endhead
    
    \midrule \multicolumn{4}{r}{{Continued on next page}} \\ \bottomrule
    \endfoot
    
    \bottomrule
    \endlastfoot
    
    Benchmark & \no & \yes & \no & \yes & \no & \yes & \yes & \yes & \no & \yes  \\ \midrule
    Open-source & \no & \yes & \no & \yes & \no & \soso & \no & \no & \no &  \yes  \\ \midrule
    OD parameters & \no & \no & \soso & \no & \no & \no & \no & \no & \yes & \yes \\ \midrule
    CFM parameters & \yes & \yes & \yes & \no & \yes & \yes & \yes & \yes & \no &  \yes \\ \midrule
    LC parameters & \yes & \no & \soso & \no & \no & \no & \no & \no & \yes & \yes  \\ \midrule
    Simulator parameters & \no & \no & \no & \soso & \soso & \no & \no & \no & \no & \yes \\ \midrule
    Hetero-geneity & \no & \no & \soso & \no & \yes & \no & \no & \no & \no &  \yes \\ \midrule
    Microscopic data & \soso & \yes & \soso & \yes & \yes & \yes & \yes & \no & \no & \yes \\ \midrule
    Aggregated data & \no & \no & \yes & \no & \yes & \no & \no & \yes & \yes & \yes  \\ \midrule
    Micro-simulator & VISSIM & None & None & None & VISSIM & None & None & AIMSUN & SUMO &  SUMO  \\ \midrule
    Setting & Urban arterial & Urban, ex-urban, rural & Varied (review paper) & Urban, varied & Urban highway, arterial, local & Varied & Rural highway & Urban \& ex-urban highway & Highway & Highway  \\ \midrule
    Source data & Per-vehicle travel times & Trajec-tories from Naples, NGSIM, Hefei, AstaZero & Varied (review paper) & Waymo Open Motion Dataset trajec-tories & 2009 cordon count program; probe vehicles & Trajec-tories from HighD, NGSIM, SPMD, Waymo, Lyft & Synthetic & Synthetic & Tenn. Dept.of Trans. \& INRIX &  I-24 INCEP-TION v1.0.0 \\ \midrule
    Data scale & 1 hr., <400 vehicles & 39 vehicle trajectories & Varied & $\sim$22mn 9- and 20-second scenarios with up to 32 agents & Counts at 45 locations; 29 probe veh. trajectories & $\sim$80K car-following events & <4 min. from one leader-follower pair & 3 hrs. of 3-min. averages at 9 detectors & 2 hrs. of edge flows and floating car speeds across $\sim$6 miles &  47 hrs. of micro. \& macro. data w/ $\sim$600K vehicles from 6.75km of two-way highway [\cite{gloudemans202324}]  \\ \midrule
    MoP & Travel time distributions & Spacing, speed, accceleration, speed std. dev. & Varied & Speed, acceleration, safety, road adherence & Counts, link avg. speed, link accel. std. dev. & Spacing for calibration; collisions also reported & Speed, spacing & Avg. speeds at detectors & Edge flows, speeds & Speed and headway distributions, agg. flows and speeds \\ \midrule
    GoF & Wilcoxon rank-sum test, Moses test, Kolmo-gorov-Smirnov test & RMSE, RMSPE, Theil's U, MAE, MAPE, RMSPE, NRMSE & Varied & Negative log likelihood & RMSE for calibration; GEH \& Theil's U also reported & MSE & RMSE, MAE, GEH statistic, Theil's U & MAE, -GEH1, MANE, RMSE, RMSENE, Theil's U & SE, AE &  See Eqn.~\ref{eq:obj_fn} \\ \midrule
    Calibration methods & GA & GA & Varied & Data-driven & GA & GA, data-driven & Downhill simplex, GA, OptQuest & SPSA\_I, SPSA\_II, OptQuest, GA, SA & Bilevel (w/ SPSA) & GA, SPSA, MA-IDM, Bilevel
    \end{longtable}
    
\end{landscape}

\section{Appendix: Benchmark Checklist}
\label{appendix:benchmark_checklist}

Below is the BetterBench checklist provided in \cite{reuel2024betterbench}, along with AutoTune's performance on each element.

\paragraph{Benchmark Design}

\begin{todolist}
    \checkeditem The tested capability, characteristic, or concept is defined
    \begin{itemize}
        \item YES: See Section~\ref{sec:problem}.
    \end{itemize}
    \checkeditem How tested capability or concept translates to benchmark task is described
    \begin{itemize}
        \item YES: See Section~\ref{sec:methodology}.
    \end{itemize}
    \checkeditem How knowing about the tested concept is helpful in the real world is described.
    \begin{itemize}
        \item YES: See Section~\ref{sec:introduction} and Table~\ref{tab:calibration_examples}.
    \end{itemize}
    \checkeditem How benchmark score should or shouldn't be interpreted/used is described
    \begin{itemize}
        \item YES: See Section~\ref{sec:methodology}.
    \end{itemize}
    \checkeditem Domain experts are involved
    \begin{itemize}
        \item YES: Authors consulted microsimulation calibration researchers and practitioners across academia, government, and industry.
    \end{itemize}
    \checkeditem Use cases and/or user personas are described
    \begin{itemize}
        \item YES: See Section~\ref{sec:methodology}.
    \end{itemize}
    \checkeditem Domain literature is integrated
    \begin{itemize}
        \item YES: See Sections~\ref{sec:related_work} and~\ref{sec:background_notation_preliminaries}.
    \end{itemize}
    \checkeditem Informed performance metric choice
    \begin{itemize}
        \item YES: See Sections~\ref{sec:related_work} and~\ref{sec:methodology}.
    \end{itemize}
    \checkeditem Metric floors and ceilings are included
    \begin{itemize}
        \item YES: See Section~\ref{sec:methodology}.
    \end{itemize}
    \xmark Human performance level is included
    \begin{itemize}
        \item NO: Manual calibration can require hundreds of human hours per scenario [\cite{alexiadis2014guidance}]. This benchmark is designed to evaluate automated calibration methods that obviate the need for this onerous labor.
    \end{itemize}
    \naitem Random performance level is included
    \begin{itemize}
        \item N/A: This applies to BetterBench's AI benchmarks, but not to microsimulation calibration. Parameters often have physical analogs so random selection is unrealistic and never done in the literature nor in practice. 
    \end{itemize}
    \checkeditem Automatic evaluation is possible and validated
    \begin{itemize}
        \item YES: The benchmark's code, which will be made available upon publication, supports this. 
    \end{itemize}
    \checkeditem Differences to related benchmarks are explained
    \begin{itemize}
        \item YES: See Section~\ref{sec:related_work}.
    \end{itemize}
    \naitem Input sensitivity is addressed
    \begin{itemize}
        \item N/A: This applies to BetterBench's AI benchmarks, but not to AutoTune. On the broader topic of random effects, the code implements random seeds and each scenario is evaluated on multiple days to account for stochastic calibration methods. 
    \end{itemize}
\end{todolist}

\paragraph{Benchmark Implementation}

\begin{todolist}
    \checkeditem The evaluation code is available
    \begin{itemize}
        \item YES: The benchmark's evaluation code will be made available upon publication.
    \end{itemize}
    \checkeditem The evaluation data or generation mechanism is accessible
    \begin{itemize}
        \item YES: The benchmark's data and evaluation code will be made available upon publication.
    \end{itemize}
    \naitem The evaluation of models via API is supported
    \begin{itemize}
        \item N/A: This applies to BetterBench's AI benchmarks, but not to microsimulation calibration.
    \end{itemize}
    \naitem The evaluation of local models is supported
    \begin{itemize}
        \item N/A: This applies to BetterBench's AI benchmarks, but not to microsimulation calibration.
    \end{itemize}
    \naitem A globally unique identifier is added or evaluation instances are encrypted
    \begin{itemize}
        \item N/A: This applies to BetterBench's AI benchmarks (where models may be trained on web data and thus risk contamination on the evaluation metrics), but not to microsimulation calibration.
    \end{itemize}
    \naitem A task to identify if model is included trained on benchmark data
    \begin{itemize}
        \item N/A: This applies to BetterBench's AI benchmarks, but not to microsimulation calibration.
    \end{itemize}
    \checkeditem A script to replicate results is explicitly included
    \begin{itemize}
        \item YES: The benchmark's code, which will be made available upon publication, supports this. 
    \end{itemize}
    \checkeditem Statistical significance or uncertainty quantification of benchmark results is reported
    \begin{itemize}
        \item YES: See Sections~\ref{sec:methodology} and~\ref{sec:results}. 
    \end{itemize}
    \naitem Need for warnings for sensitive/harmful content is assessed
    \begin{itemize}
        \item N/A: This applies to BetterBench's AI benchmarks, but not to microsimulation calibration.
    \end{itemize}
    \checkeditem A build status (or equivalent) is implemented
    \begin{itemize}
        \item YES: The benchmark's code, which will be made available upon publication, includes this. 
    \end{itemize}
    \checkeditem Release requirements are specified
    \begin{itemize}
        \item YES: The benchmark's code, which will be made available upon publication, includes this. 
    \end{itemize}
\end{todolist}

\paragraph{Benchmark Documentation}

\begin{todolist}
    \checkeditem Requirements file or equivalent is available
    \begin{itemize}
        \item YES: The benchmark's code, which will be made available upon publication, includes this. 
    \end{itemize}
    \checkeditem Quick-start guide or demo is available
    \begin{itemize}
        \item YES: The benchmark's code, which will be made available upon publication, includes a README file with necessary information. 
    \end{itemize}
    \checkeditem In-line code comments are used
    \begin{itemize}
        \item YES: The benchmark's code, which will be made available upon publication, includes this. 
    \end{itemize}
    \checkeditem Code documentation is available
    \begin{itemize}
        \item YES: The benchmark's code, which will be made available upon publication, includes this. 
    \end{itemize}
    \checkeditem Accompanying paper is accepted at peer-reviewed venue
    \begin{itemize}
        \item YES: Upon publication, this will be fulfilled. 
    \end{itemize}
    \checkeditem Benchmark construction process is documented
    \begin{itemize}
        \item YES: This is described throughout this paper and in the appendices. 
    \end{itemize}
    \checkeditem Test tasks \& rationale are documented
    \begin{itemize}
        \item YES: See Section~\ref{sec:methodology}.
    \end{itemize}
    \checkeditem Assumptions of normative properties are documented
    \begin{itemize}
        \item YES: See Section~\ref{sec:methodology}. Furthermore, users may implement custom MoP and GoF Functions to encode their desired normative properties and assess the methods upon them. 
    \end{itemize}
    \checkeditem Limitations are documented
    \begin{itemize}
        \item YES: See Section~\ref{subsec:limitations}.
    \end{itemize}
    \checkeditem Data collection, test environment design, or prompt design process is documented
    \begin{itemize}
        \item YES: See Section~\ref{sec:methodology}. For more information on the I-24 MOTION testbed and INCEPTIONv1.0.0 dataset, refer to~\cite{gloudemans202324}.
    \end{itemize}
    \checkeditem Globally unique, persistent identifier for a dataset and its metadata is provided
    \begin{itemize}
        \item YES: See~\cite{gloudemans202324}. The derived aggregate data, alongside the road network file, lane specifications, etc., will be made available upon publication, and accessible via the DOI associated with the publication. Efforts will also be made to register the derived datasets and associated metadata with their own DOI. 
    \end{itemize}
    \checkeditem Standardized metadata is included
    \begin{itemize}
        \item YES: See \cite{gloudemans202324} for information on INCEPTIONv1.0.0. The benchmark's derived data, which will be made available upon publication, uses SUMO-compatible formats or otherwise includes code to process it as such. 
    \end{itemize}
    \checkeditem Data sources and data collection process are explained
    \begin{itemize}
        \item YES: See Section~\ref{sec:methodology}.
    \end{itemize}
    \checkeditem Data preprocessing steps are described (if applicable)
    \begin{itemize}
        \item YES: See Section~\ref{sec:methodology} and Appendices~\ref{appendix:inferring_lane_boundaries}, \ref{appendix:det_counts}, and~\ref{appendix:det_noise}.
    \end{itemize}
    \naitem Data annotation process is described (if applicable)
    \begin{itemize}
        \item N/A: This applies to BetterBench's AI benchmarks, but not to microsimulation calibration.
    \end{itemize}
    \checkeditem Evaluation metric is documented
    \begin{itemize}
        \item YES: See Section~\ref{sec:methodology}.
    \end{itemize}
    \checkeditem Applicable license is specified
    \begin{itemize}
        \item YES: The benchmark's code, which will be made available upon publication, includes this. 
    \end{itemize}
    \checkeditem Data representativeness is explained (if applicable)
    \begin{itemize}
        \item YES: See Sections~\ref{sec:introduction} and~\ref{subsec:data}.
    \end{itemize}
    \checkeditem Data is documented using a standardized format
    \begin{itemize}
        \item YES: See \cite{gloudemans202324} for information on INCEPTIONv1.0.0. The benchmark's derived data, which will be made available upon publication, uses SUMO-compatible formats or otherwise includes code to process it as such. 
    \end{itemize}
\end{todolist}

\paragraph{Benchmark Maintenance}

\begin{todolist}
    \checkeditem Code usability was checked within the last year
    \begin{itemize}
        \item YES: The code has been checked within the last year. 
    \end{itemize}
    \checkeditem Maintained feedback channel for users is available
    \begin{itemize}
        \item YES: Users may email the corresponding author with feedback.
    \end{itemize}
    \checkeditem Contact person is listed
    \begin{itemize}
        \item YES: Refer to the corresponding author. 
    \end{itemize}
\end{todolist}

\section{Appendix: Traffic Microsimulation Calibrations, Applications, and Metrics}
\label{appendix:microsim_calibs_apps_metrics}

See below for a non-exhaustive list of microsimulation calibration examples and their associated measures of performance.

\begin{longtable}{|p{2.6cm}|p{3cm}|p{3cm}|p{2cm}|p{3cm}|}
    \caption{Examples of Traffic Microsimulation Calibrations with Their Applications and Metrics}
    \label{tab:calibration_examples} \\
    \toprule
    Citation & Setting & Application & Application Type & Measure of Performance \\
    \midrule
    \endfirsthead
    
    \multicolumn{5}{c}{Continued from previous page} \\
    \midrule
    Citation & Setting & Application & Application Type & Metric (Measure of Performance) \\
    \midrule
    \endhead
    
    \midrule
    \multicolumn{5}{r}{Continued on next page} \\
    \endfoot
    
    \bottomrule
    \endlastfoot
    
    \cite{boriboonsomsin2008impacts} & Highway & Comparison of high-occupancy vehicle lane configurations & Analysis, Design & GEH statistic \\
    \midrule
    \cite{cunto2008calibration} & Arterial corridor & Calibration and validation of simulated vehicle safety performance & Analysis & Crash Potential Index (a function of DRAC and MADR) \\
    \midrule 
    \cite{jung2011modeling} & Highway & Modeling safety \& operational impact of rainy weather & Planning, Analysis & Aggregate volume, speed, occupancy \\
    \midrule 
    \cite{caprani2012calibration} & Highway & Estimating bridge traffic load effect & Design & Headway \\
    \midrule
    \cite{moridpour2012enhanced} & Highway & Evaluating different lane-restriction strategies for heavy vehicles & Analysis, Design & Flows, average speeds \\
    \midrule
    \cite{gauthier2016calibration} & Highway & Evaluating the impact of projected changes to a traffic network or impacts of a construction site & Operations, Design, Planning & Time headway distributions \\ 
    \midrule
    \cite{khavas2017identifying} & Highway & Modeling traffic in inclement weather & Planning, Operations & Capacity, speed at free-flow and at capacity  \\
    \midrule
    \cite{amirjamshidi2019multi} & Urban network of arterials, local
    roads, \& freeways  & Modeling carbon emissions & Analysis & Counts, link average speeds, acceleration standard deviation \\
    \midrule
    \cite{chen2019assessing} & Urban expressway & Assessing the influence of adverse weather using a driving simulator & Analysis & Link average speeds \\
    \midrule 
    \cite{ard2020microsimulation} & Highway & Assessing the potential impact of mixed heterogeneous platoons & Future Concepts & Time headway distributions, travel time \\
    \midrule
    \cite{pan2021evaluation} & Intersection & Assessing Continuous Flow Intersection design and performance & Design & Capacity \\
    \midrule 
    \cite{ren2024traffic} & Highway & Assessing the potential impact of mixed heterogeneous platoons & Future Concepts &  \\
    \midrule
    \cite{qu2025revisiting} & Highway & Assessing the correlation between simulated and field-observed safety conflicts & Analysis & Counts, trajectory positions and speeds \\

    \bottomrule
\end{longtable}

\section{Appendix: Inferring Lane Boundaries}
\label{appendix:inferring_lane_boundaries}

\begin{figure}[t]
    \centering \includegraphics[width=0.95\textwidth]{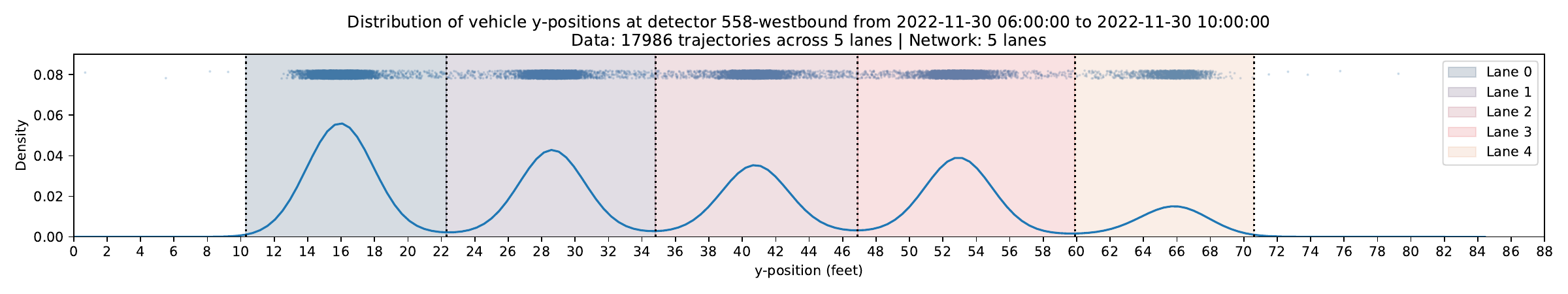} 
    \caption{Example of kernel density estimate plot and associated lane boundaries inferred from latitudinal position data at one virtual detector. Adapted from [\cite{hickert2026autotune}].}
    \label{fig:lane_boundaries_558wb} 
\end{figure}

\begin{figure}[ht]
    \centering \includegraphics[width=0.95\textwidth]{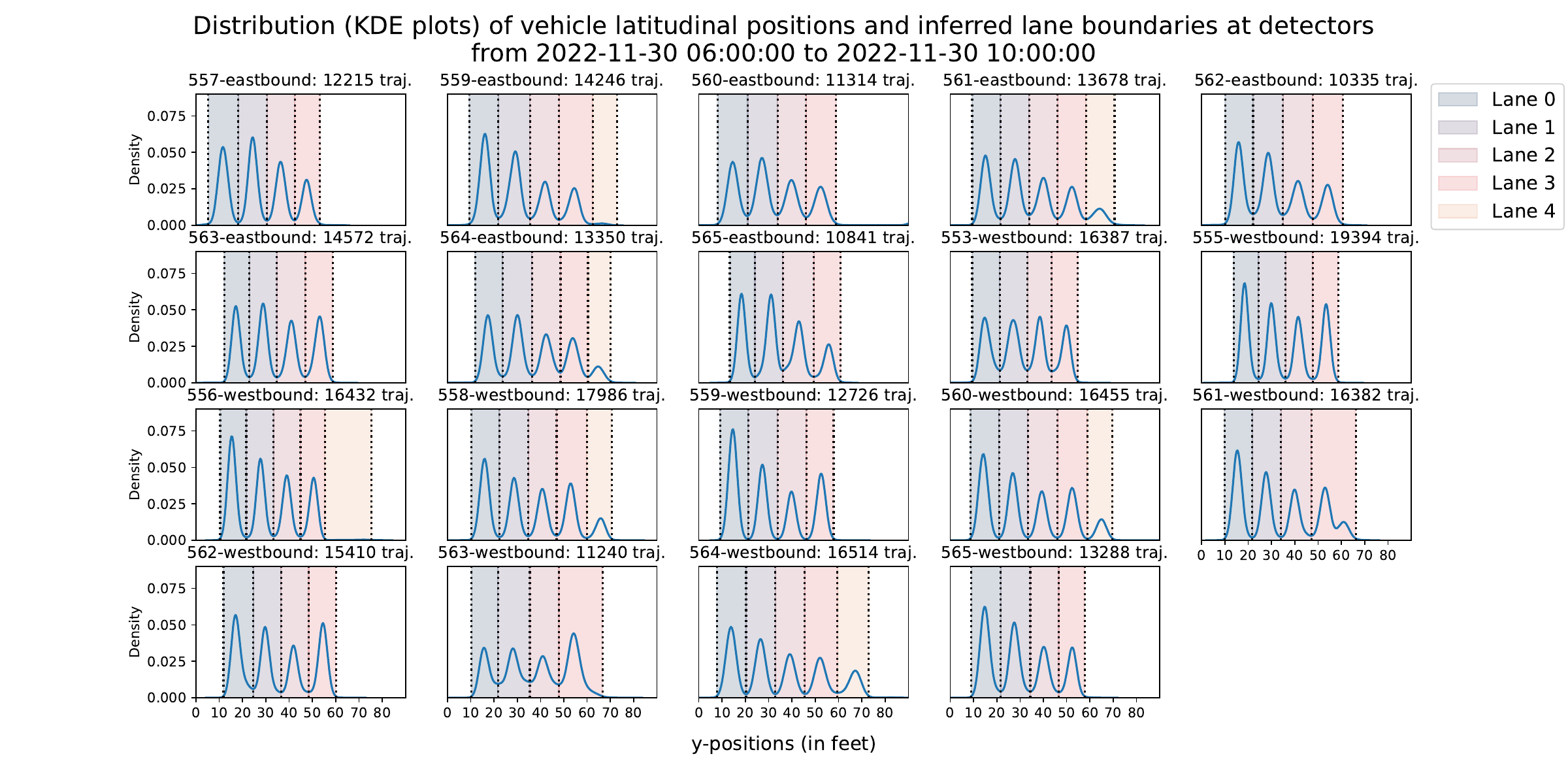} 
    \caption{Kernel density estimate plots and associated lane boundaries inferred from latitudinal position data at each virtual detector.}
    \label{fig:lane_boundaries_all} 
\end{figure}

The I-24 MOTION testbed features 20 Tennessee Department of Transportation (TDoT) radar detection systems (RDS) distributed across the corridor. However, because RDS logs are unavailable for the specific period covered by the INCEPTIONv1.0.0 dataset, we created `virtual' detectors at those exact physical locations by extracting macroscopic flow and velocity metrics directly from the trajectory data, aggregating readings into 30-second bins to mirror standard highway sensor outputs.

An obstacle encountered during this extraction was the absence of explicit lane assignments -- an issue frequently observed in aggregated highway traffic datasets [\cite{Caltrans2023}]. To resolve this, we implemented a data-driven lane inference method, utilizing kernel density estimation (KDE) on the lateral vehicle coordinates at each detector site to identify local minima as lane dividers. See Figures~\ref{fig:lane_boundaries_558wb} and \ref{fig:lane_boundaries_all}. Inferred boundaries were visually inspected against the KDE plots and adjusted or omitted in a small number of cases with nearby ramps, sparse outer-lane detections, or lane-splitting artifacts.
 
Of course, as future INCEPTION data updates reduce the error in the data, the benchmark can incorporate those for enhanced accuracy in its assessments.

\section{Appendix: Converting INCEPTION data to detector counts}
\label{appendix:det_counts}

INCEPTION trajectories are converted to loop-detector-like measurements using virtual detectors placed at the same milemarker locations as the SUMO E1 detectors. For each aggregation interval $\Delta t$ (30 s in the benchmark data), trajectories are filtered by direction and detector location, and a vehicle is counted if the trajectory point closest to the detector falls within the half-open interval $[t, t+\Delta t)$. The output count $qPKW$ is therefore a lane-level vehicle count per aggregation interval, not an hourly flow rate.

Lane assignment is performed at the detector crossing point using inferred lane-boundary locations. Trajectories whose crossing point falls outside the outermost inferred lane boundaries are excluded. Because the INCEPTION lane convention and SUMO network lane convention use opposite ordering, lane indices are remapped when writing the detector-level CSVs so that each row matches the corresponding SUMO detector ID.

Speed is estimated longitudinally over a short centered time window around the detector crossing, rather than from an instantaneous derivative. Specifically, the distance traveled in the longitudinal coordinate is divided by the elapsed time between samples before and after the detector crossing, then converted to km/hr. This smoothing step is used to reduce sensitivity to known differentiation error in computer vision-derived trajectories, where first and second derivatives can amplify localization noise [\cite{coifman2017critical}]. The output speed is the mean of the vehicle speeds assigned to that detector lane and aggregation interval; intervals with no detected vehicles are assigned zero count and zero speed.

The preprocessing script also records quality-control diagnostics, including the number of candidate trajectories, the number excluded by time-window and lane-boundary checks, and the maximum longitudinal distance between the selected trajectory sample and the detector location.

\section{Appendix: Detector Noise}
\label{appendix:det_noise}

Sensor error data is extracted from the plots in \cite{middleton2000initial}, which evaluated the performance of various detectors for freeway applications. We selected the 3M microloop as the sensor from which to construct noise distributions because the study found it to be the most consistent overall and of the technologies studied most closely resembles PeMS detectors. The microloop sensor records both speeds and counts, so we handled the noising process for each independently. Although the original source data was not published in the study, we extracted it manually across eight days using the study's figures and extraction tools from \cite{WebPlotDigitizer}.

From the extracted empirical speed error distribution, speed errors were sampled uniformly at random with replacement and applied to the original per-lane speed data at each 30-second increment. 

This same strategy could not be directly applied to count data, due to the integer nature of counts and their low values. Attempting to do so would underestimate the desired error due to rounding. Instead, from the empirical distribution we sampled count error at 15-minute, per-detector intervals, in line with the microloop's count measurement aggregation frequency. We then applied this error percentage at the same granularity to the extracted I-24 data. We then distributed the resulting integer difference between the noisy 15-minute sums and the originals among the 30-second counts, applying them in a random weighted fashion to each 30-second data point, where weights are based on the original proportions of each 30-second count within its 15-minute sum. This method was compared to other redistribution schemes and found to provide the best empirical match to the source error data. 

\section{Appendix: Calibration method parameters and ranges}
\label{appendix:cal_params_ranges}

See below for each method's specific calibration parameters, their descriptions, ranges, and initialization values.

\begin{table}[h!]
\centering
\caption{Parameters for SUMO Baseline}
\begin{tabular}{|p{2cm}|p{2.2cm}|p{5cm}|p{2cm}|p{1.8cm}|} 
\hline
\textbf{Parameter Category} & \textbf{Parameter} & \textbf{Description} & \textbf{Calibration range} & \textbf{Starting value (if applicable)} \\
\hline
OD & OD Matrix Entries & Vehicle origins, destinations, and departure times & [0-6000] & 3000 \\
\hline
\end{tabular}
\label{tab:method_name_here}
\end{table}

\begin{table}[h!]
\centering
\caption{Parameters for MA-IDM}
\begin{tabular}{|p{2cm}|p{2.2cm}|p{5cm}|p{2cm}|} 
\hline
\textbf{Parameter Category} & \textbf{Parameter} & \textbf{Description} & \textbf{Calibration range} \\
\hline
CFM & Desired speed ($v_0$) & The maximum speed a driver wants to travel at under free-flow conditions (i.e., no leader ahead). & (0, $+\infty$)\\
\hline
CFM & Jam distance ($s_0$) & The minimum gap maintained at standstill (e.g., at a red light or in a jam). & (0, $+\infty$)\\
\hline
CFM & Desired time headway ($T$) & The time gap the driver wants to maintain behind a leading vehicle. & (0, $+\infty$)\\
\hline
CFM & Maximum acceleration ($a$) & The maximum acceleration the driver is willing to apply. & (0, $+\infty$)\\
\hline
CFM & Comfortable deceleration ($b$) & The maximum comfortable deceleration the driver applies to avoid collisions. & (0, $+\infty$)\\
\hline
CFM & Acceleration exponent ($\delta$) & Controls the smoothness of acceleration when approaching the desired speed. & 4 (fixed)\\
\hline
CFM & Lengthscale ($\ell$) & A hyperparameter of the GP that determines how quickly the influence of past actions decays over time. & (0, $+\infty$)\\
\hline
CFM & Variance ($\sigma^2$) & Another GP hyperparameter representing the variance of the residuals, capturing the uncertainty in the driver's acceleration behavior not explained by the IDM. & (0, $+\infty$)\\
\hline
\end{tabular}
\label{tab:MA-IDM-params}
\end{table}

\begin{table}[h!]
    \centering
    \caption{Parameters for Genetic Algorithm and SPSA}
    \begin{tabular}{|p{2cm}|p{2.7cm}|p{5.8cm}|p{2cm}|} 
    \hline
    \textbf{Parameter Category} & \textbf{Parameter} & \textbf{Description} & \textbf{Calibration range} \\
    \hline
    CFM & Minimum Headway ($\tau$) & Desired minimum time headway behind a leading vehicle. & [0.2, 3] \\
    \hline
    CFM & EmergencyDecel & Maximum deceleration to avoid an accident. & [6, 12] \\
    \hline
    CFM & accel & Maximum acceleration the driver applies. & [0.5, 5] \\
    \hline
    CFM & decel & Maximum comfortable deceleration the driver applies. & [1, 6] \\
    \hline
    CFM & minGap & Minimum gap maintained at standstill (bumper-to-bumper). & [0.2, 5] \\
    \hline
    CFM & Acceleration exponent ($\delta$) & Controls smoothness of acceleration when approaching desired speed. & [2, 6] \\
    \hline
    CFM & actionStepLength & Duration between vehicle decisions for speed and lane-changing. & [0.9, 1.1] \\
    \hline
    CFM & speedFactor & Factor scaling a vehicle's desired maximum speed. & [0.9, 1.1] \\
    \hline
    LC & lcStrategic & Eagerness for long-term, route-driven lane changes. & [0, 1] \\
    \hline
    LC & lcCooperative & Willingness to adjust speed to facilitate other vehicles' lane changes. & [0, 1] \\
    \hline
    LC & lcAssertive & Willingness to accept smaller gaps during lane changes. & [0, 1] \\
    \hline
    LC & lcSpeedGain & Eagerness for lane changes to gain speed. & [0, 1] \\
    \hline
    LC & lcKeepRight & Tendency to keep to the rightmost available lane. & [0, 1] \\
    \hline
    OD & ($taz_x$, $taz_y$) & Flow range for vehicles between origin-destination (TAZ) pairs. & [0, 2000] \\
    \hline
    HET & variance & Controls variance in driver behavior or attributes. & [0.02, 1] \\
    \hline
    HET & speedFactor Variance & Variance in speedFactor among different drivers. & [0.02, 0.05] \\
    \hline
    SIM & numChunks & (Project-specific) Division of simulation or data processing into chunks. & [1, 2, 4, 6] \\
    \hline
    SIM & warmupTime & Initial simulation period for network to reach stable traffic. & [0, 300] \\
    \hline
    SIM & stepLength & Duration of a single simulation step in seconds. & [0.1, 0.25, 0.5, 1] \\
    \hline
    \end{tabular}
    \label{tab:GA_SPSA_params}
\end{table}


\begin{small}
\begin{sloppypar} 
\bibliographystyle{authordate1} 

\setlength{\bibsep}{0pt}

\bibliography{References}

\end{sloppypar}
\end{small}


\end{document}